\documentclass[a4paper,fleqn,usenatbib,useAMS]{mnras}

\usepackage{graphicx}	
\usepackage{amsmath}	
\usepackage{amssymb}	
\usepackage{multicol}        
\usepackage{bm}		
\usepackage{pdflscape}	

\usepackage[T1]{fontenc}
\usepackage{ae,aecompl}

\usepackage{newtxtext,newtxmath}

\title[MBM 110]{The Sejong Open cluster Survey (SOS) VI. A small star-forming
region in the high Galactic latitude molecular cloud MBM 110}

\author[Sung, Bessell \& Song]{Hwankyung Sung$^{1}$\thanks{Contact e-mail: {sungh@sejong.ac.kr}}, Michael S. Bessell$^{2}$, Inseok Song$^{3}$
\\
$^{1}$Department of Physics and Astronomy, Sejong University,
    209 Neungdong-ro, Gwangjin-gu, Seoul 05006, Korea \\
$^{2}$Research School of Astronomy \& Astrophysics, The Australian
    National University, Canberra, ACT 2611, Australia \\
$^{3}$Department of Physics and Astronomy, University of Georgia,
    Athens, GA 30602, USA}

\date{Last updated 2019 March 25}

\pubyear{2019}

\begin{document}
\label{firstpage}
\pagerange{\pageref{firstpage}--\pageref{lastpage}}
\maketitle

\begin{abstract}
We present optical photometric, spectroscopic data for the stars in the high
Galactic latitude molecular cloud MBM 110. For the complete membership
selection of MBM 110, we also analyze {\it WISE} mid-infrared data and
{\it Gaia} astrometric data. Membership of individual stars is critically
evaluated using the data mentioned above. The {\it Gaia} parallax of
stars in MBM 110 is 2.667 $\pm$ 0.095 mas ($d = 375 \pm 13 pc$), which confirms
that MBM 110 is a small star-forming region in the Orion-Eridanus superbubble.
The age of MBM 110 is between 1.9 Myr and 3.1 Myr depending on the adopted
pre-main sequence evolution model. The total stellar mass of MBM 110 is
between 16 $M_\odot$ (members only) and 23 $M_\odot$ (including probable
members). The star formation efficiency is estimated to be about 1.4\%.
We discuss the importance of such small star formation regions in
the context of the global star formation rate and suggest that a galaxy's
star formation rate calculated from the H$\alpha$ luminosity may underestimate
the actual star formation rate. We also confirm a young brown
dwarf member based on photometry, spectroscopy, and astrometry.
\end{abstract}




\newpage

\section{Introduction}

MBM 110, also known as L1634 \citep{Lyn62}, is one of the high Galactic latitude
molecular clouds discovered by \citet{mbm85}, and is one of a dozen  cometary
clouds  in the Orion-Eridanus superbubble \citep{Bal08}. Star formation is
ongoing in many of these clouds \citep{acl08}. Most investigators considered
that the massive stars in the Orion OB association might have triggered star
formation in these regions \citep{lc07,acl08}. MBM 110 is about $3.^\circ 8$
to the west of the Orion Nebula, and is  facing towards the Ori OB1 association.
\citet{mmm86} suggested that MBM 110 may be either remnants of the molecular
cloud material from which the nearby OB association formed, or clouds pushed
to their current location by pressure associated with energetic events
accompanying the evolution of the OB association. Interestingly, MBM 110 does
not have a conspicuous X-ray emitting young population \citep{san95}.

Herbig-Haro objects, bipolar outflows and the red nebulous object, RNO 40 in
L1634 have attracted several detailed studies \citep{bpt93,beh02,ocsd04}.
\citet{beh02} studied the embedded young stellar objects (YSOs) IRAS 05173-0555
and IRS 7 based on multicolour observations from centimeter to sub-millimeter
wavelengths, and confirmed the Class 0 nature of IRAS 05173-0555. Several
H$\alpha$ emission stars are known from H$\alpha$ surveys \citep{Ste86,kiso}.
However, a detailed study for the young star content or star formation history
has not yet been performed.

In this, the sixth paper of the Sejong Open cluster Survey (SOS) project 
\citep{slb13}, we present photometric and spectroscopic data for the stars
in the high Galactic latitude molecular cloud MBM 110. In Section 2, we
describe the photometric, spectroscopic, and astrometric data for MBM 110.
The membership of suspected members is assigned in the same section.
The Hertzsprung-Russell diagram (HRD) for the members and probable members
is constructed in Section 3. The mass and age of individual stars are
determined in the HRD based on two popular pre-main sequence (PMS)
evolution models. In addition, the star formation efficiency of MBM 110
is also determined in the section. The importance of small star-forming
regions (SFRs) like MBM 110 is discussed in Section 4. The summary of
this investigation is presented in Section 5.

\begin{table*}
 \caption{Observation Log}
 \label{tab_log}
 \begin{tabular}{c|l|c|l}
  \hline
   Telescope & Date of Obs. & detector & Filter \& Exposure Time \\
  \hline
SSO 1m & 2003. 10. 20  & WFI & B: 15${}^s$, 600${}^s$,  V: 5${}^s$, 240${}^s$, R: 5${}^s$, 180${}^s$,  H$\alpha$: 45${}^s$, 900${}^s$, I: 5${}^s$, 90${}^s$  \\
 & 2003. 10. 23 & WFI &B: 7${}^s$, 60${}^s$, 1200${}^s$, V: 5${}^s$, 45${}^s$,  900${}^s$  \\
& & &R: 5${}^s$, 45${}^s$, 900${}^s$,  H$\alpha$: 30${}^s$, 120${}^s$, 600${}^s \times 3$,  I: 5${}^s$, 30${}^s$, 600${}^s$  \\
Maidanak 1.5m & 2007. 1. 25 & SNUCam & U: 30${}^s$, 600${}^s$, B: 7${}^s$, 600${}^s$, V:  5${}^s$, 300${}^s$, R: 5${}^s$, 180${}^s$, I: 5${}^s$, 60${}^s$ \\
 & 2007. 1. 26 & SNUCam & U: 30${}^s$, 600${}^s$, B: 7${}^s$, 600${}^s$, V:  5${}^s$, 300${}^s$, H$\alpha$: 30${}^s$, 600${}^s$, I: 5${}^s$, 60${}^s$ \\ \hline\hline
Telescope & Date of Obs. & spectrograph & grating \& spectral resolution  \\ \hline
Lick 3m & 2004. 2. 13 & KAST & Blue: 600g/mm ($1.86 \AA$/pix), Red: 830g/mm ($1.69 \AA$/pix)  \\
BOAO 1.8m & 2004. 11. 5 \& 6 & BOLS & 1200g/mm ($0.66 \AA$/pix @ H$\alpha$)  \\
                    & 2004. 11. 7 & BOLS & 1200g/mm ($0.67 \AA$/pix @ H$\beta$) \\
                    & 2010. 12. 11  & BOLS & 500g/mm ($2.02 \AA$/pix @ H$\alpha$)  \\
                    & 2011. 12. 12 \& 13  & BOLS & 500g/mm ($2.02 \AA$/pix @ H$\alpha$)  \\
SSO 2.3m & 2012. 11. 1--5 & WiFeS & Blue: 708g/mm, Red: 398g/mm ($1.0 \AA$/pix)  \\
                  & 2012. 11. 12 -- 13 & WiFeS & Blue: 708g/mm, Red: 398g/mm ($1.0 \AA$/pix)   \\
  \hline
 \end{tabular}
\end{table*}

\section{Observations and membership selection}

The number of near-infrared excess stars found in MBM 110 attracted an optical
imaging study of the region. We performed optical photometry with the wide
field imager (WFI) of the 1m telescope at Siding Spring Observatory (SSO)
for a comprehensive study of YSOs in the region. In January 2007, we obtained
additional optical images of the central region with the AZT-22 1.5m telescope
at Maidanak Astronomical Observatory (MAO), Uzbekistan. Based on the photometry,
we selected targets for spectroscopic observations - H$\alpha$ emission stars
and stars in the PMS locus. For a complete census of YSOs in MBM 110 we aimed
to obtain medium resolution spectra for the stars with H$\alpha$ emission
as well as for stars in the PMS locus using several telescopes - Lick 3m
telescope, Bohyun-san Optical Astronomy Observatory (BOAO) 1.8m telescope,
and SSO 2.3m telescope. The log of observations is presented in Table
\ref{tab_log}.

As optical photometry cannot provide a complete census of YSOs with
circumstellar disks, we downloaded {\it WISE} mid-infrared (MIR) data and
searched for Class I and II YSOs in the region. In addition, as weak-line
T Tauri stars (TTSs) may be missed in our H$\alpha$ survey, we searched
for MIR excess emission stars and obtained their optical spectra.
We also searched for kinematic members using the {\it Gaia} Data Release
2 (DR2) astrometric data.

\begin{figure*}
   \includegraphics[width=2.0\columnwidth]{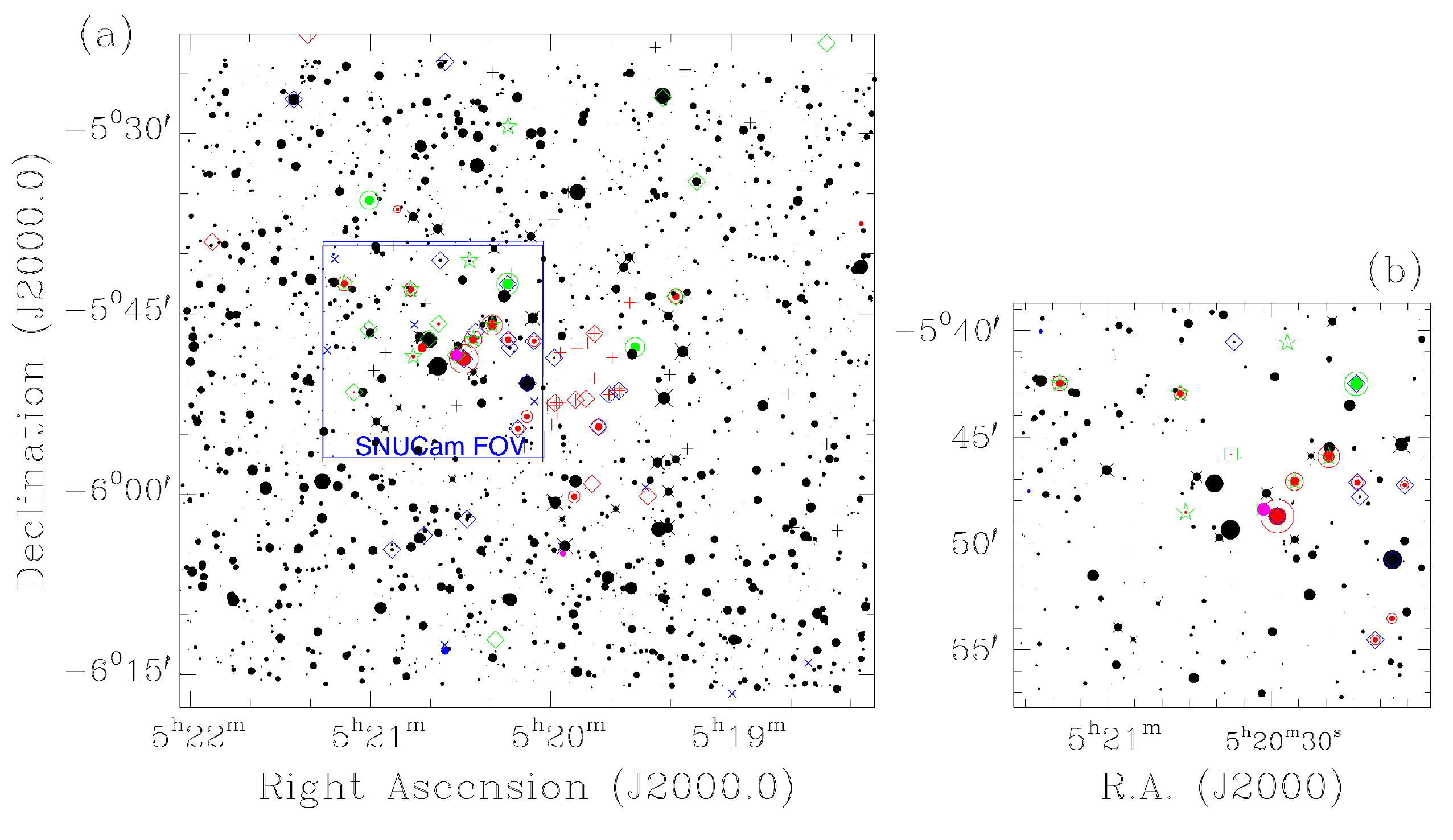}
     \caption{(a) Finder chart for the stars brighter than $I$ = 17 mag
     from SSO WFI observations and (b) for the stars brighter than $V$ = 19
     mag from MAO SNUCam observations. The size of the dot is proportional
     to the brightness of the star (I-band magnitude  for (a) and V-band
     for (b)). Blue crosses, red plus symbols, and black plus symbols denote
     stellar or non-stellar X-ray emission objects, Herbig-Haro objects,
     and IR sources, respectively. Colour of dots represents the type of
     membership - red: H$\alpha$ emission star from photometry or     
     previous H$\alpha$ surveys, blue: X-ray emission star, magenta:
     X-ray emission star with H$\alpha$ emission, green: stars
     with strong Li absorption, and black: the other stars. We added
     an additional symbol - red circle: spectroscopically confirmed
     H$\alpha$ emission star, green circle: star with strong Li absorption,
     black cross: spectroscopically confirmed star with no H$\alpha$ emission,
     red diamond: Class I YSO, green diamond: Class II YSO, green star:
     Class II proper motion member, blue diamond: proper motion member.}
     \label{map} 
\end{figure*}

\subsection{Optical Photometry}
\subsubsection{SSO 1m WFI Observations}

The observations were conducted on two nights on 2003 October 20 \& 23 with
the $8k \times 8k$ mosaic CCD camera WFI [field of view (FOV) $\approx 52'
\times 52'$] of the 1m telescope at SSO.  The exposure times and filters
used in the observation are detailed in Table \ref{tab_log}. All the
preprocessing was carried out using the IRAF\footnote{Image Reduction and
Analysis Facility is developed and distributed by the National Optical
Astronomy Observatories, which is operated by the Association of Universities
for Research in Astronomy under the cooperative agreement with the National
Science Foundation.}/MSCRED package. Instrumental magnitudes were obtained
using the IRAF version of DAOPHOT \citep{Ste91} via point spread function (PSF)
fitting. The atmospheric extinction coefficients and transformation
coefficients used were those in \citet{sbc08}, but the photometric zero-points
of magnitudes and colours were adjusted to those of the Maidanak 4k CCD data.
Figure \ref{map} (left) shows the finder chart based on the WFI observations.
We also marked stellar or non-stellar X-ray emission objects, Herbig-Haro
objects, and IR sources in the figure. The information on these sources was
obtained from the astronomical database {\it Simbad}
\footnote{\url{http://simbad.u-strasbg.fr/simbad/}}.

Photometric data for 10246 stars were obtained from the SSO WFI observations.
The only saturated star in the field is HD 34909 (K0III, $V$ = 8.00).
From the ($V-I$, $R-$H$\alpha$) and  ($R-I$, $R-$H$\alpha$) two-colour
diagrams (TCDs) we selected 13 H$\alpha$ emission stars and 2 H$\alpha$
emission candidates (see Table \ref{member}). However, 6 of 13 H$\alpha$
emission stars and 1 of 2 H$\alpha$ emission candidates were previously
known H$\alpha$ emission stars from other H$\alpha$ surveys \citep{Ste86,kiso}.
The photometric data for 2052 stars brighter than $I$ = 17 mag are
presented in Table \ref{wfi_data} (Photometric data for fainter stars are
available from H.S.). We do not present 
the colour-magnitude diagrams (CMDs) and TCDs from SSO WFI observations 
to avoid the repetition of similar figures. The main purpose of the SSO WFI
observations was to find the H$\alpha$ emission stars in MBM 110 as well
as to isolate probable members with no appreciable H$\alpha$ emission for
the spectroscopic observations. For the target selection for spectroscopic
observations, we identified the locus of H$\alpha$ emission stars in the
($I, ~V-I$) CMD. The PMS locus for the spectroscopic observations was
slightly wider than the two isochrones in Figure \ref{m4k_cmd} (upper left).

\subsubsection{MAO SNUCam Observations \label{mao_phot}}

For the detection of ultraviolet (UV) excess emission among TTSs in MBM 110,
$UBVRI$ and H$\alpha$ images for the central region were obtained on 2007
January 25 \& 26 at MAO with the AZT-22 (1.5m) telescope (f/7.74) and
a thinned Fairchild 486 CCD (SNUCam - \citealt{ikc10}). The exposure times
and filters used in the observation are listed in Table \ref{tab_log}.
All the preprocessing required to remove the instrumental signature was
done using the IRAF/CCDRED package. Instrumental magnitudes were obtained
using IRAF/DAOPHOT via PSF fitting for the target images and via simple
aperture photometry for standard stars.
Details of the transformations to the standard system can be found in \citet{lsb09}. The atmospheric extinction coefficients
and photometric zero-points derived from the standard star photometry are presented in Table \ref{mao_coef}.
The finder chart based on MAO observations is shown in Figure \ref{map} (right). Table \ref{m4k_data} is
the photometric data for 798 stars from the MAO observations.

\begin{figure*}
\includegraphics[width=1.8\columnwidth]{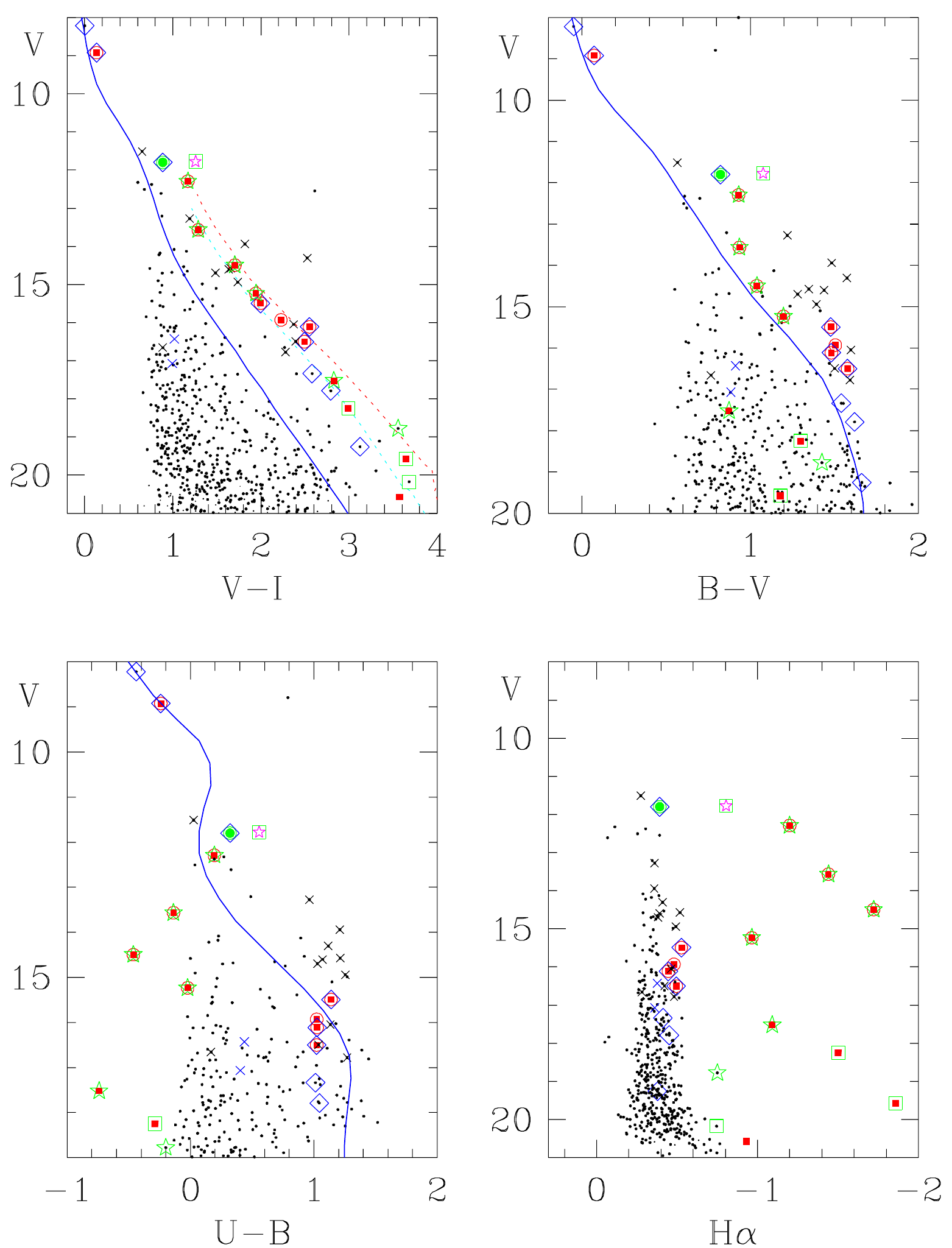}
\caption{Colour-magnitude diagrams of MBM 110 from MAO SNUCam observations.
$UBV$ photometric data for the two brightest stars were obtained from the  $UBV$ database of \citet{mer91} in {\it VizieR}.
The blue solid line represents the ZAMS relation for a distance modulus of  7.87 mag and $E(B-V)$ = 0.11 mag.
A red square, red triangle, blue cross, magenta star mark, and green dot represents an H$\alpha$ emission star,
H$\alpha$ emission candidate, X-ray emission star, X-ray emission star with H$\alpha$ emission, and star
with appreciable Li absorption, respectively. We also added an additional symbol - a circle, black cross, red square, green square,
green star mark, and blue diamond for the spectroscopically confirmed member, spectroscopically non-member,
Class I object, Class II object, Class II and proper motion member, and proper motion member, respectively.
The green and red dashed lines in the upper left panel represent, respectively,  5 Myr  and 2 Myr
isochrone from \citet{bhac15}.}
\label{m4k_cmd} 
\end{figure*}

Figure \ref{m4k_cmd} shows the CMDs from the MAO observations. In the ($V, ~ V-I$) CMD
young PMS stars (H$\alpha$ emission stars, Class II stars and most of the proper motion members) are well
aligned to the PMS isochrone (\citealt{bhac15}, hereafter BHAC) of age 2 Myr (red) and 5 Myr (green). However, there are several
non-member stars (black crosses) in or near the PMS locus. And two  proper motion members are below the isochrone of age 5 Myr.
In the ($V, ~ B-V$) and ($V, ~ U-B$) CMDs many PMS stars are bluer than the zero-age main sequence (ZAMS).
However, in the ($V, ~ U-B$) CMD, three bright members ($V$ = 11.5 -- 12.5 mag), four H$\alpha$ emission stars and
5 proper motion members
(three of them are H$\alpha$ emission stars) are near the ZAMS. These stars are either PMS members with no appreciable
mass accretion or non-members with strong chromospheric activity. The H$\alpha$ index in Figure \ref{m4k_cmd} (lower right) is defined in \citet{scb00}. 

\begin{table}
 \caption{Atmospheric Extinction Coefficients and Photometric Zero Points}
 \label{mao_coef}
 \begin{tabular}{c@{}||c@{}|cccc}
  \hline
Date & Filter & $k_1$ & $k_2$ & $C_0$ & $\zeta$ \\
  \hline
              & $I$    & 0.028 $\pm$ 0.001 & ...               & $R-I$ & 23.199 $\pm$ 0.013 \\
              & $I$    &                   &                   & $V-I$ & 23.201 $\pm$ 0.013 \\
 2007.   & $R$    & 0.076 $\pm$ 0.004 & ...               & $R-I$ & 23.593 $\pm$ 0.014 \\
Jan. 25 & $V$    & 0.121 $\pm$ 0.006 & ...               & $V-I$ & 23.642 $\pm$ 0.011 \\
              & $V$    &                   &                   & $B-V$ & 23.641 $\pm$ 0.009 \\
              & $B$    & 0.238 $\pm$ 0.002 & 0.033 $\pm$ 0.002 & $B-V$ & 23.491 $\pm$ 0.005 \\
              & $U$    & 0.417 $\pm$ 0.009 & 0.018 $\pm$ 0.001 & $U-B$ & 21.705 $\pm$ 0.008 \\ \hline
              & $I$    & 0.039 $\pm$ 0.005 & ...               & $R-I$ & 23.218 $\pm$ 0.003 \\
              & $I$    &                   &                   & $V-I$ & 23.220 $\pm$ 0.003 \\
2007.  & H$\alpha$& 0.076 $\pm$ 0.014 & ...               & $V-I$ & 19.615 $\pm$ 0.017 \\
Jan. 26 & $V$    & 0.131 $\pm$ 0.003 & ...               & $V-I$ & 23.665 $\pm$ 0.004 \\
              & $V$    &                   &                   & $B-V$ & 23.660 $\pm$ 0.003 \\
              & $B$    & 0.227 $\pm$ 0.003 & 0.034 $\pm$ 0.004 & $B-V$ & 23.476 $\pm$ 0.002 \\
              & $U$    & 0.414 $\pm$ 0.026 & 0.033 $\pm$ 0.001 & $U-B$ & 21.679 $\pm$ 0.003 \\
  \hline
\end{tabular}
\end{table}

\begin{figure*}
\includegraphics[width=2\columnwidth]{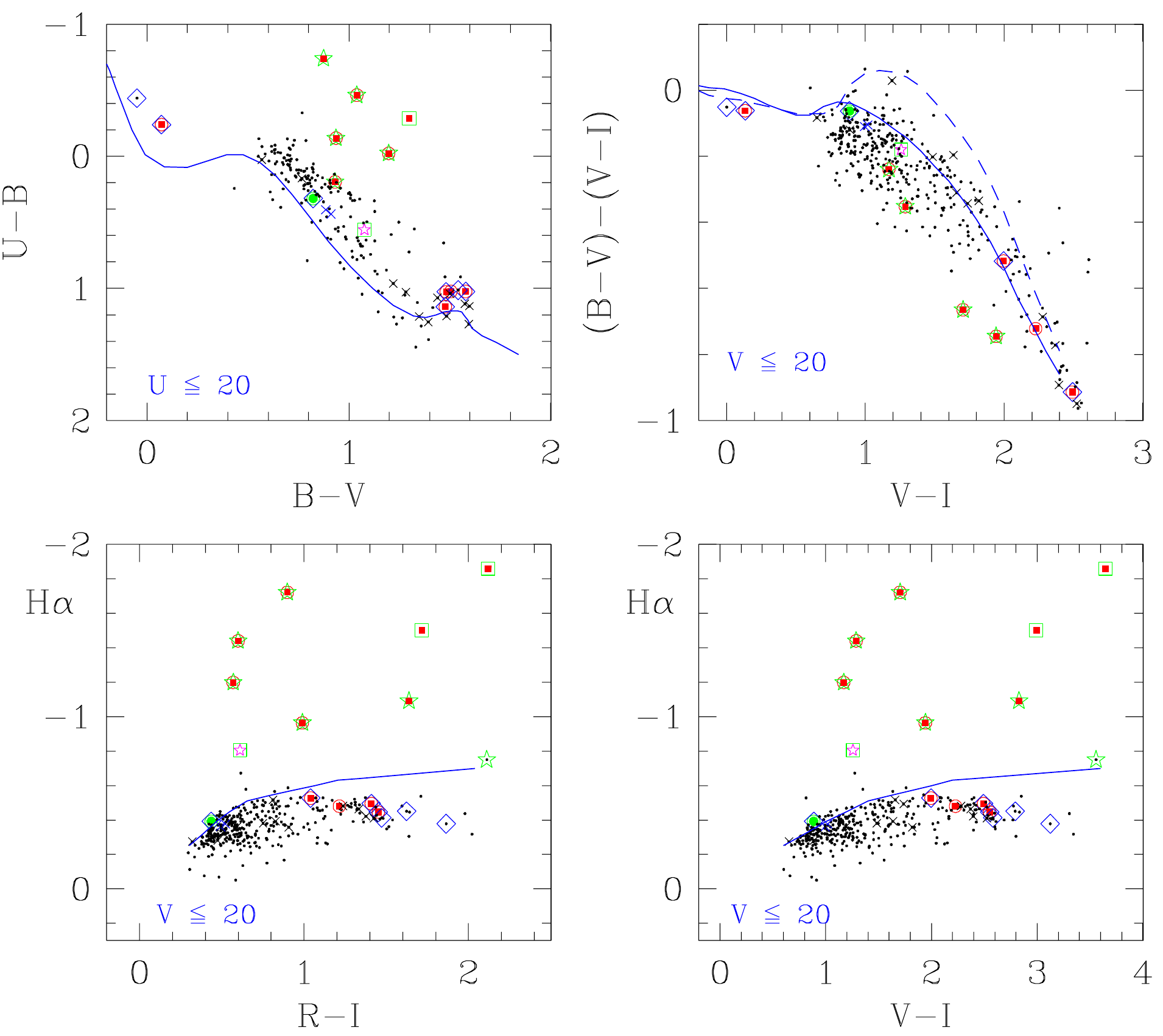}
\caption{Two-colour diagrams of MBM 110 from MAO SNUCam observations.
The blue solid line in the upper panel represents the intrinsic colour-colour relation of main sequence (MS) stars \citep{slb13},
while the dashed line (upper right) is that  of giant stars. The TCDs in the lower panel show
the selection criterion for H$\alpha$ emission stars. The solid line in the lower panel represents the photospheric
level of unreddened normal stars.
All symbols are the same as in Figure \ref{m4k_cmd}}
\label{m4k_ccd}
\end{figure*}

Figure \ref{m4k_ccd} shows the TCDs from the MAO observations. The most massive star in the WFI FOV, HD 34835, is
slightly reddened [$E(B-V)$ = 0.11]. The peculiar star HD 34890 (= HIP 24930) is a single line spectroscopic binary [see
Figure \ref{fig_spec} (left)] with an unseen companion, and so is not used in the reddening determination.
We also determined the colour excesses in $V-I$, $V-J$, $V-H$, and $V-K_s$ for HD 34385, and derived
the total-to-selective extinction ratio $R_V$ using the colour-excess ratios (see \citealt{slb13,sbc17}).
Although the $E(B-V)$ of HD 34385 is small and so the colour excess ratios may have large errors, the $R_V$'s from
these ratios are consistent within 1.4 times of the standard deviation and give $3.42 \pm 0.10$.
The most prominent feature in the $UBV$ as well as the $BVI$ TCD is the strong $U$ and $B$ excess of 5 low-mass PMS members.
However, as mentioned in the previous paragraph, 4 weak H$\alpha$ emission stars do not show any appreciable UV excess.
The lower panels in Figure \ref{m4k_ccd} show the selection criteria for the H$\alpha$ emission stars from H$\alpha$
photometry. The selection of
H$\alpha$ emission stars from H$\alpha$
photometry is limited to classical T Tauri stars (CTTSs) (see Table \ref{member} or Figure 6 of \citealt{sbc08}).
From the TCDs we selected 8 H$\alpha$ emission stars and 1 H$\alpha$ emission candidate. There are 3 stars (black dots) 
between the H$\alpha$ emission stars and the fiducial line. They were not selected as H$\alpha$ emission stars 

\begin{landscape}
 \begin{table}
  \caption{Photometric data from SSO 1m WFI observation$^{*}$}
  \label{wfi_data}
  \begin{tabular}{ccccccccccccccccccccl}
  \hline
   ID & $\alpha_{\rm J2000}$ & $\delta_{\rm J2000}$ & $I$ & $R-I$ &
   $V-I$ & $B-V$ & $R$-H$\alpha$ & $\epsilon_{I}$ & $\epsilon_{R-I}$ &
   $\epsilon_{V-I}$ & $\epsilon_{B-V}$ & $\epsilon_{R-{\rm H}\alpha}$ &
   \multicolumn{5}{c}{N$_{\rm obs}$} & 2MASS ID$^{a}$ & M$^b$ & Remark \\
  \hline
WFI 1026 &  5:20:13.70 &-5:47:48.6& 14.787 &  1.456 &  2.700 &  1.508 & -3.291 &  0.048 &  0.038 &  0.045 &  0.001 &  0.002 & 5& 5& 3& 3& 4&05201379-0547487&  & PM member   \\
WFI 1027 &  5:20:13.78 &-6:11:15.6& 16.152 &  0.429 &  0.879 &  0.797 & -3.301 &  0.001 &  0.000 &  0.002 &  0.006 &  0.011 & 5& 5& 5& 3& 3&05201374-0611160&  &   \\
WFI 1028 &  5:20:13.92 &-5:45:53.2& 16.405 &  1.401 &  2.626 &  1.550 & -3.229 &  0.018 &  0.045 &  0.072 &  0.092 &  0.025 & 5& 3& 3& 2& 2&05201403-0545533&  &   \\
WFI 1029 &  5:20:14.03 &-6:05:20.8& 14.111 &  0.393 &  0.786 &  0.719 & -3.383 &  0.001 &  0.014 &  0.021 &  0.007 &  0.011 & 5& 5& 5& 5& 5&05201400-0605213&  &   \\
WFI 1030 &  5:20:14.14 &-5:47:08.4& 13.444 &  0.990 &  1.997 &  1.482 & -3.280 &  0.010 &  0.029 &  0.072 &  0.005 &  0.001 & 5& 5& 5& 4& 5&05201423-0547085&S &  PM member  \\
WFI 1031 &  5:20:14.26 &-6:08:41.1& 10.736 &  0.359 &  0.813 &  0.855 & -3.327 &  0.002 &  0.015 &  0.018 &  0.007 &  0.002 & 4& 3& 3& 3& 3&05201422-0608416&  &   \\
WFI 1032 &  5:20:14.28 &-5:29:26.0& 16.282 &  1.978 &  3.570 &   ...  & -2.817 &  0.007 &  0.002 &  0.027 &   ...  &  0.000 & 5& 3& 2& 0& 2&05201430-0529259&H & Class II, PM member  \\
WFI 1033 &  5:20:14.36 &-5:42:30.2& 10.899 &  0.425 &  0.863 &  0.792 & -3.293 &  0.018 &  0.002 &  0.001 &  0.001 &  0.006 & 3& 3& 3& 3& 3&05201443-0542301&L & PM member  \\
WFI 1034 &  5:20:14.66 &-5:32:48.4& 16.232 &  0.625 &  1.234 &  1.124 & -3.292 &  0.017 &  0.018 &  0.007 &  0.001 &  0.007 & 5& 5& 4& 2& 3&05201468-0532485&  &   \\
WFI 1035 &  5:20:14.73 &-5:38:09.3& 15.659 &  0.553 &  1.048 &  0.848 & -3.348 &  0.004 &  0.017 &  0.014 &  0.006 &  0.007 & 5& 5& 5& 3& 4&05201474-0538092&  &   \\
  \hline
  \end{tabular}
  $^{*}$ Table \ref{wfi_data} is presented in its entirety in the electronic edition of the MNRAS. A portion is shown here for guidance regarding its form and content. Units of right ascension are hours, minutes, and seconds of time, and units of declination are degrees, arcminutes, and arcseconds. \\
  $^{a}$ A, B, or b are added at the end of 2MASS ID if two or more stars are matched with a 2MASS source within a matching radius of 1$''$. A or B: The bright or faint component of a 2MASS source whose $I$ magnitude difference is less than 1 mag. b: The faint component of a 2MASS source whose $I$ magnitude difference is greater than 1 mag. \\
  $^{b}$ membership - S: H$\alpha$ emission star confirmed from spectroscopy, X: X-ray emission star, H: H$\alpha$ emission star, h: H$\alpha$ emission candidate, +: X + H, -: X + h, n: star with no H$\alpha$ emission from spectroscopy \\
\end{table}

 \begin{table}
  \caption{Photometric data from MAO observations$^{*}$}
  \label{m4k_data}
  \begin{tabular}{cccccccccccccccccccccc@{}ccl}
  \hline
   ID & $\alpha_{\rm J2000}$ & $\delta_{\rm J2000}$ & $V$ &
   $R-I$ & $V-I$ & $B-V$ & $U-B$ & $R$-H$\alpha$ & $\epsilon_{V}$ &
   $\epsilon_{R-I}$ & $\epsilon_{V-I}$ & $\epsilon_{B-V}$ & $\epsilon_{U-B}$ &
   $\epsilon_{R-{\rm H\alpha}}$ & \multicolumn{6}{c}{N$_{\rm obs}$} & 2MASS ID &
   WFI ID & M$^{a}$ & $\rm W(H\alpha)$  \\
  \hline
M4k405 &  5:20:45.70 &  -5:48:32.3 &  17.519 &   1.638 &   2.828 &   0.874 &  -0.740 &  -1.091 &   0.055 &   0.014 &   0.051 &   0.004 &   0.012 &   0.006 & 4 & 2 & 4 & 4 & 4 & 2 & 05204569-0548323 & 1314 & H, II, PM &       \\
M4k406 &  5:20:45.83 &  -5:48:55.5 &  20.261 &   0.614 &   1.273 &   1.125 &  ...  &  ...  &   0.079 &   0.079 &   0.097 &   0.065 &  ...  &  ...  & 2 & 1 & 2 & 2 & 0 & 0 &                  & 8682 &   &          \\
M4k407 &  5:20:45.89 &  -5:52:57.6 &  21.726 &  ...  &   2.147 &  ...  &  ...  &  ...  &   0.164 &  ...  &   0.184 &  ...  &  ...  &  ...  & 1 & 0 & 1 & 0 & 0 & 0 &                  &      &      &       \\
M4k408 &  5:20:46.22 &  -5:50:00.5 &  20.868 &   1.300 &   2.526 &  ...  &  ...  &  -0.381 &   0.015 &   0.045 &   0.032 &  ...  &  ...  &   0.164 & 2 & 1 & 2 & 0 & 0 & 1 & 05204621-0550004 & 8691 &      &       \\
M4k409 &  5:20:46.27 &  -5:39:01.7 &  18.577 &   0.409 &   0.819 &   0.696 &  -0.010 &  ...  &   0.001 &   0.007 &   0.005 &   0.014 &   0.040 &  ...  & 2 & 2 & 2 & 1 & 1 & 0 &                  & 8693 &      &       \\
M4k410 &  5:20:46.33 &  -5:47:39.6 &  17.212 &   0.663 &   1.298 &   0.958 &   0.241 &  -0.268 &   0.014 &   0.015 &   0.000 &   0.014 &   0.006 &   0.015 & 4 & 2 & 4 & 4 & 3 & 2 & 05204632-0547397 & 1320 &      &       \\
M4k411 &  5:20:46.43 &  -5:40:26.7 &  16.132 &   0.512 &   1.015 &   0.853 &   0.303 &  -0.340 &   0.001 &   0.010 &   0.003 &   0.009 &   0.015 &   0.017 & 4 & 2 & 4 & 4 & 4 & 2 & 05204641-0540267 & 1322 &      &       \\
M4k412 &  5:20:46.47 &  -5:40:41.0 &  18.549 &   0.576 &   1.080 &   0.911 &   0.233 &  -0.346 &   0.005 &   0.011 &   0.003 &   0.012 &   0.076 &   0.039 & 4 & 2 & 4 & 2 & 2 & 1 & 05204648-0540409 & 8700 &      &       \\
M4k413 &  5:20:46.56 &  -5:55:51.9 &  18.120 &   0.615 &   1.275 &   1.129 &   1.047 &  -0.451 &   0.004 &   0.022 &   0.005 &   0.006 &   0.082 &   0.010 & 4 & 2 & 4 & 3 & 2 & 2 & 05204655-0555519 & 1325 &      &       \\
M4k414 &  5:20:46.60 &  -5:53:38.3 &  18.933 &   1.480 &   2.557 &   1.610 &  ...  &  -0.409 &   0.003 &   0.016 &   0.009 &   0.008 &  ...  &   0.024 & 3 & 2 & 3 & 2 & 0 & 1 & 05204659-0553382 & 1326 &   PM &       \\
M4k415 &  5:20:46.68 &  -5:42:57.7 &  15.234 &   0.990 &   1.942 &   1.197 &  -0.023 &  -0.964 &   0.046 &   0.013 &   0.035 &   0.017 &   0.031 &   0.015 & 4 & 2 & 4 & 4 & 4 & 2 & 05204667-0542577 & 1327 & H, II, PM & -35.4, -32.4 \\
\hline
\end{tabular}
$^{*}$ Table \ref{m4k_data} is presented in its entirety in the electronic edition of the MNRAS.
A portion is shown here for guidance regarding its form and content. Units of right ascension are
hours, minutes, and seconds of time, and units of declination are degrees, arcminutes, and arcseconds. \\
$^{a}$membership - X: X-ray emission star, x: X-ray emission candidate, H: H$\alpha$ emission star, h: H$\alpha$ emission candidate, +: X + H, -: X + h , II: YSO Class II object, PM: proper motion member \\
\end{table}
\end{landscape}

\noindent
because they are all very faint ($V \geq 19$) and the errors of their H$\alpha$ index are about 0.1 mag.

\subsection{Spectroscopy}

\begin{figure*}
   \includegraphics[width=1.8\columnwidth]{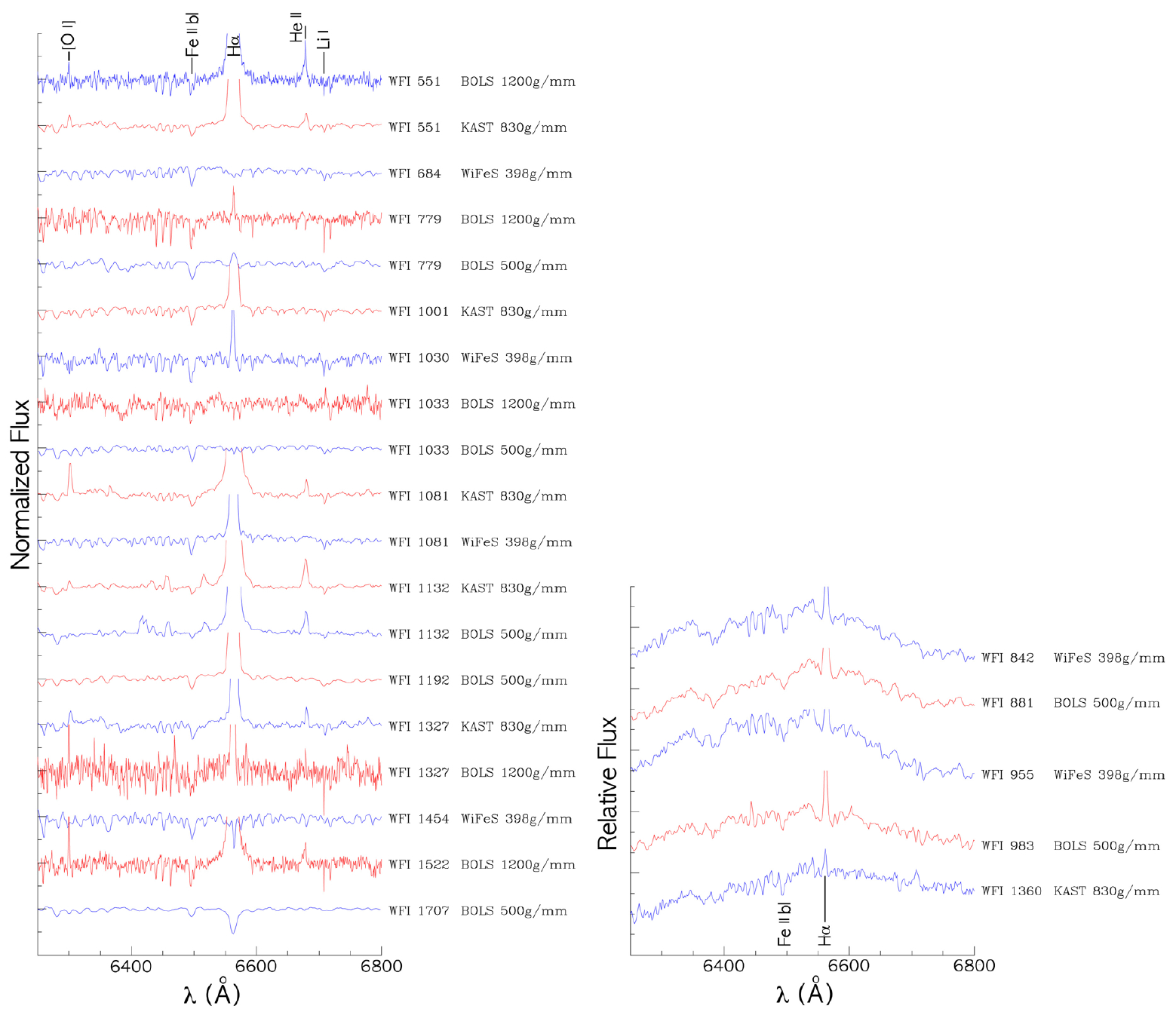}
     \caption{ Red spectrum of stars with H$\alpha$ emission and/or Li {\small I} $\lambda 6708$ absorption.
     (a) K- or early-type stars, (b) M-type stars}
     \label{fig_tts}
\end{figure*}

As H$\alpha$ photometry can only detect stars with strong H$\alpha$ emission ($W_{\rm H\alpha} \gtrsim 20\AA$,
i.e. CTTS - \citealt{sbc08}) and we detected only a dozen H$\alpha$ emission stars from the SSO WFI observations,
we decided to conduct spectroscopic observations for a complete census of PMS stars in MBM 110.
On 2004 February 12, we obtained the spectra for 12 stars including HD 34835 and HD 34890
with the dual beam spectrograph KAST of the Lick 3m Shane Reflector. We confirm H$\alpha$ emission in 6 stars
(WFI 551, 1001, 1081, 1132, 1327 \& 1360 - see Figure \ref{fig_tts}). On 2004 November 5 -- 8,
we observed stars in the PMS locus using the long slit mode of BOES (BOLS) (grating: 1200g/mm).
H$\alpha$ emission and Li {\small I} $\lambda 6708\AA$ absorption
are confirmed for WFI 551, 779, 1327, and 1522. However, although the H$\alpha$ line of WFI 1033 is in absorption,
Li {\small I} $\lambda 6708\AA$ absorption is relatively strong. WFI 1132 shows a strong emission at H$\beta$ ($W_{\rm 
H\beta} = -31.4\AA$).

For the efficient detection of faint H$\alpha$ emission stars, we used the 500g/mm grating of the BOLS
in 2010 December and 2011 December observing runs. H$\alpha$ emission was found in WFI 779, 881, and 1192 among the 6 stars observed in the 2010 December run and in WFI 1132 among 8 stars observed in the 2011 December run.
Among 7 stars without
H$\alpha$ emission in the 2011 December run, WFI 1033 showed an appreciable absorption in Li {\small I} $\lambda 6708\AA$.
In 2012 November we observed 22 stars in the PMS locus with the integral field spectrograph WiFeS of the SSO 2.3m
telescope, and confirmed H$\alpha$ emission for 5 stars (WFI 842,
899, 955, 1030 \& 1081). Two stars (WFI 684 \& 1454) showed Li {\small I} $\lambda 6708\AA$ absorption.
The spectra around H$\alpha$ and Li {\small I} $\lambda 6708\AA$ are shown in Figure \ref{fig_tts}.
We could not detect any appreciable H$\alpha$ emission and Li {\small I} $\lambda 6708\AA$ absorption for 33 stars.
These stars are marked ``n'' in the 20th column in Table \ref{wfi_data} and in the 24th column
of Table \ref{m4k_data}.

\begin{figure}
   \includegraphics[width=\columnwidth]{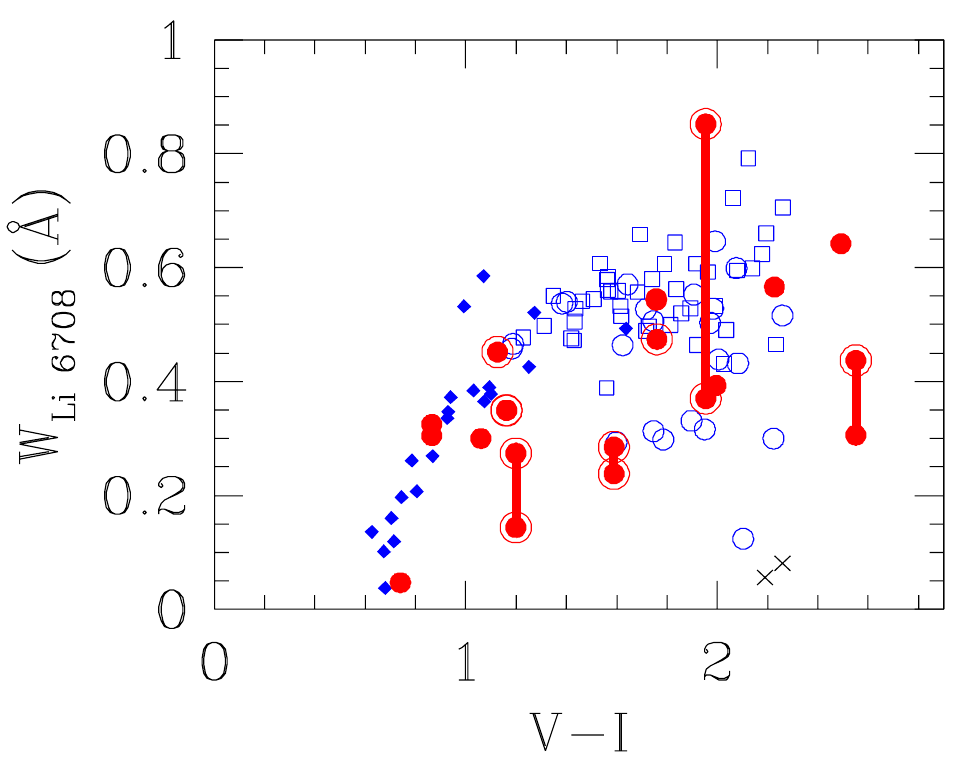}
     \caption{Equivalent width of Li $\lambda 6708\AA$ as a function of $V-I$.
       Red dots represent stars in MBM 110. We added an additional circle for CTTSs in MBM 110.
       Two crosses denote the maximum value of $\lambda 6708\AA$ absorption for two non-member stars.
       The line joining two dots indicates different values from different observations for a given star.
       Blue circles, blue squares, and blue diamonds represent, respectively,
       H$\alpha$ emission stars, stars with H$\alpha$ in absorption, and no information on H$\alpha$ emission or absorption
       in the young open cluster NGC 2264 for comparison \citep{lsk16}.}
   \label{fig_Li}
\end{figure}

At the estimated age of MBM 110 (see section \ref{age}),
we do not expect any M-stars to have significantly depleted Li. Furthermore, the Li $\lambda 6708\AA$ feature is a highly saturated
spectral feature such that a visible sign of depleted Li requires more than 90\% depletion in M-type stars.
Therefore, true MBM 110 M-type members are expected to show a strong (i.e. nearly undepleted) Li $\lambda 6708\AA$ feature.
Figure \ref{fig_Li} shows the equivalent width of Li $\lambda 6708\AA$ ($W_{\rm Li 6708}$) as a function
of $V-I$. For comparison, the $W_{\rm Li 6708}$ data for the young open cluster NGC 2264 (age: about 3 Myr
- \citealt{sb10}) from \citet{lsk16} are plotted in the figure.  The variation in Li absorption strength is most 
pronounced for M-type stars and is, therefore, a good age indicator for young stars. The Li strength is slightly
weaker than that of stars in the young open cluster NGC 2264, and is very similar to that of young stars in the
$\eta$ Cha ($W_{\rm Li 6708}$ = 0.4 -- 0.6 $\AA$ for M0 -- M3 stars) \citep{zs04}.

HD 34835 has the earliest type B5III/IV ({\it Simbad})  in the observed field.  The source of the spectral type is \citealt{hs99}.
The spectrum of this star was obtained twice
- once at Lick and once at BOAO. The line ratio between He {\small I} $\lambda 4471 \AA$ and Mg {\small II} $\lambda 4482 \AA$
indicates a spectral type of B6. The line ratio between He {\small I} $\lambda 4009 \AA$ and He {\small I} $\lambda 4026\AA$ also
supports this. In addition, the strength of C II $\lambda 4267\AA$  relative to the Balmer lines and the He {\small I} lines is comparable
to that of spectral type B5 -- B7. We classify the spectral type of this star as B6.

\begin{figure*}
   \includegraphics[width=1.8\columnwidth]{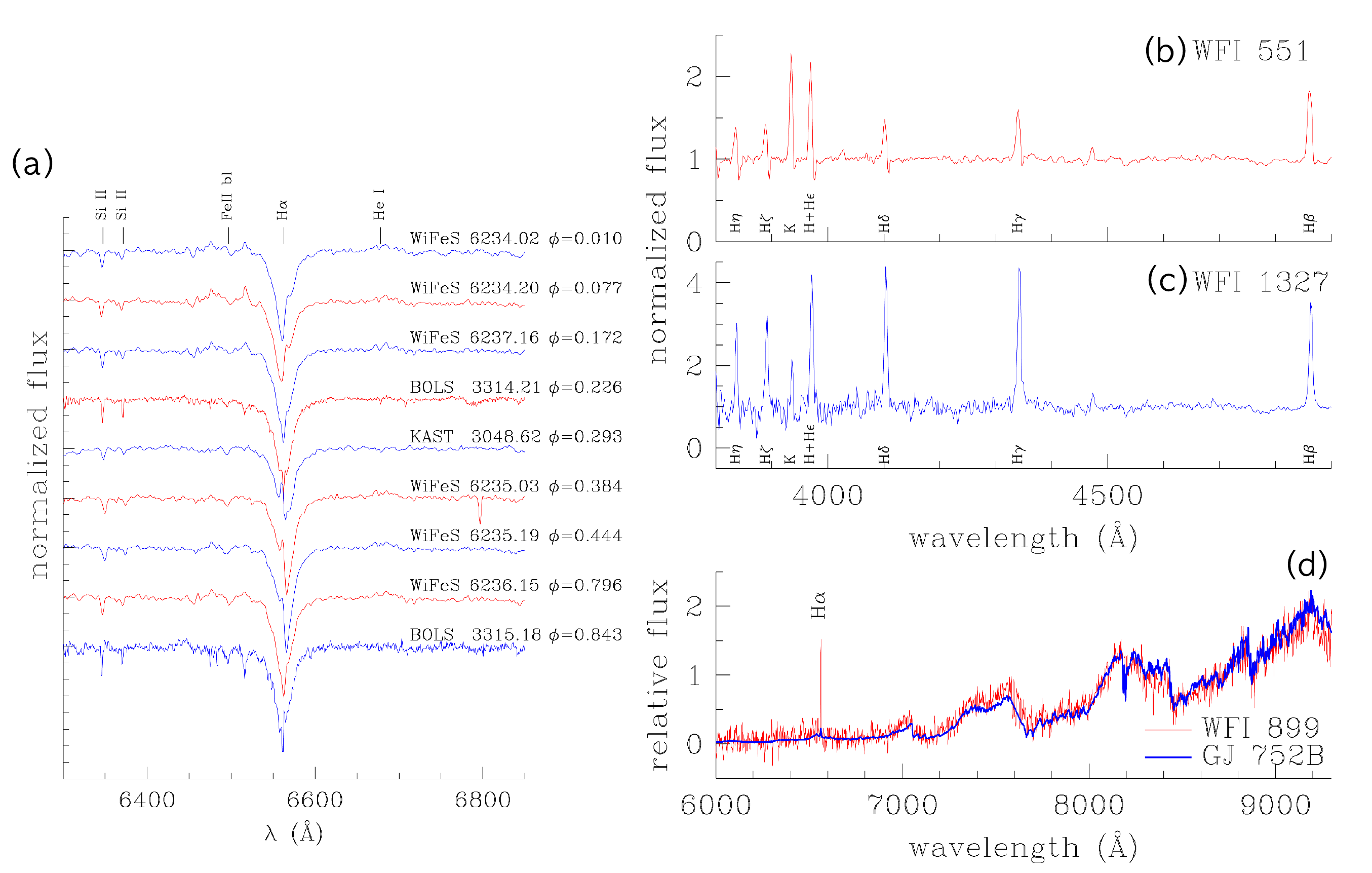}
     \caption{Spectrum of special targets. (a) H$\alpha$ line profile variation of HD 34890,
      (b) WFI 551, (c) WFI 1327, (d) WFI 899 (thin red line) and
      M8V star GJ 752B (=VB 10, thick blue line) }
   \label{fig_spec}
\end{figure*}

The spectral type of HD 34890 was A0pec initially, but is currently B9IV/V in {\it Simbad}.
This star is a single line spectroscopic binary with an unseen companion. More detailed spectroscopic characteristics of
this star will be presented in a forthcoming paper. The H$\alpha$ line of HD 34890 [see
Figure \ref{fig_spec} (a)] shows an emission component within a strong absorption line. Both line centre and
strength vary with time. The period of variation is about 2.705 days.
In the spectrum of HD 34890, He {\small I} lines are weak, but definitely seen, indicating a late-B type star.
The He {\small I} $\lambda 4471 \AA$ and Mg {\small II} $\lambda 4482 \AA$  ratio is about 1/2, which indicates a spectral type
of B8 -- B9. He {\small I} $\lambda 4009\AA$ is very weak or absent, and Ca {\small II} $\lambda 4267\AA$ is invisible.
These features imply that HD 34890 is a B9 star. The spectral type of HD 34890 is adopted as B9.

WFI 1081 is a known H$\alpha$ emission star (StH$\alpha$ 37). The spectrum of this star was obtained twice - once at Lick
and once at SSO. In the Lick/KAST spectrum all Balmer lines are very strong [$W(\rm H\alpha) = -167\AA$]. Ca {\small II} H \& K lines
show strong emission, but are weaker than the Balmer lines. [O {\small I}] $\lambda 6300\AA$, He {\small I} $\lambda\lambda$ 5876,
6678 \& 7065$\AA$ are in emission. However, in the SSO/WiFeS spectra, the Ca {\small II} H \& K lines are stronger than
most Balmer lines except H$\alpha$. We could not find any emission at [O {\small I}]$\lambda 6300\AA$ and He {\small I} $\lambda 6678\AA$.
He {\small I} $\lambda\lambda$ 5876 \& 7065$\AA$ lines are in emission, but weaker that those of the Lick/KAST spectra. In addition,
the Ca {\small II} IR triplet ($\lambda\lambda$ 8498, 8542, \& 8662$\AA$) shows strong emission.

WFI 551 is a known H$\alpha$ emission star (Kiso A-0975 47). H$\alpha$ emission is detected in both H$\alpha$ photometry
and spectroscopy. The spectrum of the star was obtained twice (once with Lick/KAST, the other with BOLS/g1200). The equivalent
width of H$\alpha$ shows a large variation ($-63.4 \AA$ at Lick and $-107.5\AA$ at BOAO). In addition, He I $\lambda\lambda$
5875, 6678, 7065$\AA$, and [O I] $\lambda 6300\AA$ also show emission. Interestingly, all Balmer lines
(H$\gamma$, H$\delta$, H$\epsilon$, H$\zeta$, and H$\eta$) and Ca {\small II} K in the blue show inverse P-Cygni profiles
[see the upper panel of Figure \ref{fig_spec} (b)], indicating WFI 551 is a PMS star with active mass accretion.
The Li absorption strength (two connected points at $V-I \approx 1.6$ in Figure \ref{fig_Li}) is weaker than 
that of young stars in NGC 2264, and is caused by a veiling effect due to strong mass accretion. {\it WISE} MIR colours
also indicate the existence of  a circumstellar disk around the star (Class II). Its {\it Gaia} proper motion also indicates the membership
of the star. However, the parallax is larger than that of MBM 110 members. 

WFI 1327 is an H$\alpha$ emission star, Kiso A-0975 57. The H$\alpha$ emission of this star is confirmed from
MAO observations as well as spectroscopic observations at Lick and BOAO. However, $R-$H$\alpha$ from
SSO WFI observation is not pronounced. The spectrum of WFI 1327 shows strong emission in the Balmer lines.
Interestingly, the strength of H$\gamma$ and H$\delta$ are stronger than H$\beta$ in the Lick/KAST spectrum
[see Figure \ref{fig_spec} (b) (middle)]. Many He {\small I} and [O {\small I}] $\lambda6300\AA$ lines show
emission. The Li absorption strength also showed a large difference between the two observations as seen in Figure \ref{fig_Li}
(two connected points at $V-I \approx 2.0$).

WFI 899 is one of the faintest stars in the PMS locus and is suspected to be a young brown dwarf.
The spectrum of WFI 899 was obtained on 2012 November 4 with WiFeS at SSO.
H$\alpha$ emission is well pronounced. However, although the absorption of
Li $\lambda 6708\AA$ is suspected, it cannot be confirmed due to the weak signal. The spectrum is shown
in the lower panel of Figure \ref{fig_spec} (b). The spectrum is compared with the well-known M8V GJ 752B (=VB 10).
The two spectra are quite similar, but there are some features that are weaker (Na {\small I}, K {\small I}, and CaH) indicating
that WFI 899 has a lower gravity than GJ 752B. WFI 899 is the only confirmed young brown dwarf in MBM 110.
The emission of H$\alpha$ with a higher resolution spectrum (R = 7000) was confirmed on 2012 November 24.
Although the errors are somewhat large, Gaia DR2 parallax as well as proper motions support the membership of WFI 899.

WFI 1132 is a well-known CTTS (V534 Ori, StH$\alpha$ 38). The spectra of the star were obtained three times (Lick/KAST,
BOLS/g1200 @H$\beta$, \& BOLS/g500). H$\alpha$ emission is very strong [W(H$\alpha$) $\approx 100\AA$]. H$\beta$ is
also very strong, but is weaker than the Ca {\small II} K line. He {\small I} $\lambda\lambda$4026, 4388, 4472, 4922, 5016, 5876,
6678, \& 7065$\AA$ and [O {\small I}] $\lambda 6300\AA$ are in emission. The H$\beta$ emission line obtained with the BOLS/g1200
shows a double peak (or weak self-absorption) profile. 
The Li absorption strength (two connected points at $V-I \approx 1.2$ in Figure \ref{fig_Li}) is weaker than 
that of young stars in NGC 2264, and is caused by a veiling effect due to strong mass accretion.

\subsection{{\it Gaia} Astrometric Data}

\begin{figure*}
  \includegraphics[width=1.8\columnwidth]{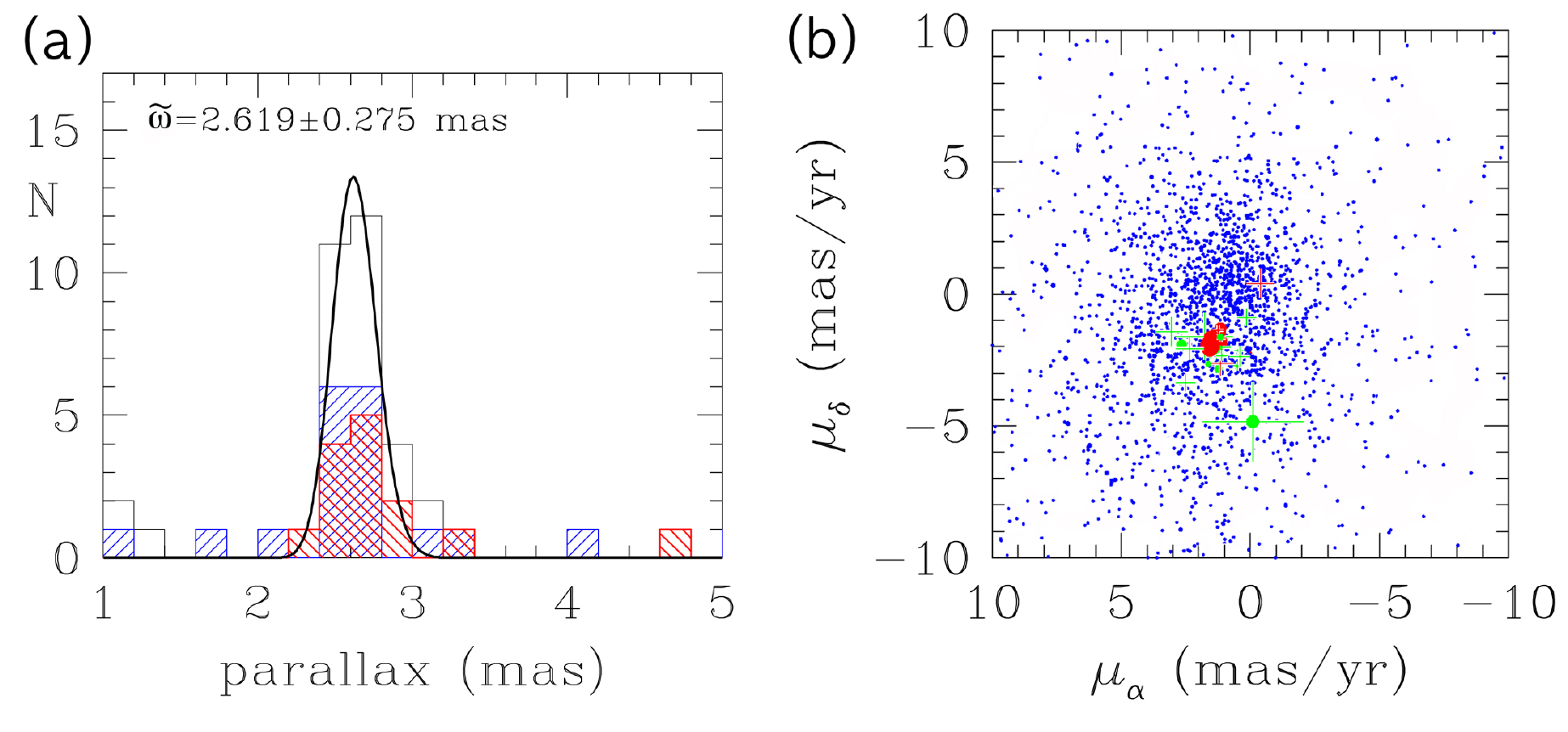}
     \caption{(a) Gaia DR2 parallax distribution of probable members of MBM 110. Blue, red hatched histograms and open
     histograms represent the parallax distribution of spectroscopically confirmed young stars, Class II objects, and all probable
     members, respectively. (b) Proper motion of stars in MBM 110. 
     The proper motion of members (red) and probable members (green) are marked with error bars. 
     Small dots represent field stars.}
   \label{fig_gaia}
\end{figure*}

Recently, the astrometric satellite {\it Gaia} of the European Space Agency released 5-dimensional
data for about 1.7 billion sources.
MBM 110 is a loose SFR with a dozen known member stars and only one early-type member in the MS band, so
for the distance determination and for the selection of members not detected from H$\alpha$ photometry or
MIR excess emission, we downloaded {\it Gaia} DR2 data from {\it VizieR}\footnote{\url{http://vizier.u-strasbg.fr}} for $1^\circ \times 1^\circ$ FOV centred on the A0pec star HD 34890. 
To determine the distance of MBM 110, we checked the distribution of parallaxes of probable member stars.
Figure \ref{fig_gaia} (a) shows parallax distribution of spectroscopically confirmed members (blue histogram - H$\alpha$
emission and Li {\small I} $\lambda 6708$ absorption), that of Class II YSOs (red histogram), and all probable members.
Although the parallaxes of probable member stars spread over a large range, many are concentrated between 2.4 mas and 2.8 mas.
We tried fitting the histogram with a Gaussian function, and found that the parallax of probable members of MBM 110 is
2.619 $\pm$ 0.275 mas (equivalently $d = 382 \pm 40 pc$). And the median and the robust scatter estimate
[RSE - see the footnote 4 of \citet{lhb18} for the definition of RSE] are 2.614 mas and 0.095 mas ($d = 383 \pm 14 pc$), respectively.

The member stars selected from the optical study are distributed over $35'$ in diameter, which is equivalent to about $3.9 pc$.
This value is consistent with the expanded size due to a random internal velocity dispersion of about $1 km ~s^{-1}$
for about 4 Myr (see section \ref{age}). However, the radial scale from {\it Gaia} DR2 data appeared to be large
probably due to the uncertainty in the current {\it Gaia} DR2 parallaxes.
The distance to MBM 110 is very similar to the recent distance of the Orion Nebula Cluster ($\tilde{\omega}$ = 2.530 $\pm$ 0.001 mas,
$d = 389 \pm 3 pc$) from the 6-dimensional analysis 
\citep{kcs18}.

As members of star clusters or stellar groups formed from a single cloud, they share common kinematic
properties inherited from their natal cloud. Proper motions and radial velocities should, therefore, be good
membership criteria. We checked the proper motion of stars in the FOV and that of member stars. Unfortunately,
most stars in the FOV and the member stars occupy nearly the same region in the proper motion plane [see
Figure \ref{fig_gaia} (b)]. This fact implies that proper motion is not a good membership selection criterion
for MBM 110. The proper motion of definite members (red dots with error bar) and of probable members (green
dots with error bar) is also shown in the figure. As can be seen in the figure, the proper motion error of probable members
and one member is much larger than that of most member stars.

In the {\it Gaia} DR2 astrometry paper,  \citet{lhb18} identifies a global zero-point shift of 0.029mas.
In addition, the zero-point residual varies from region to region on small and large scales (see Fig. 12 -- 15 of \citealt{lhb18}).
To check the zero-point near MBM 110, we downloaded the astrometric data for quasar candidates within 1 degree
from HD 34890. There are 36 AllWISE AGN sources \citep{sdd15} and 1 prototype ICRF3 VLBI source. Among them,
3 objects have only two-parameter solutions.
The median value and RSE of the parallaxes are -0.053 mas and 0.438 mas, respectively. 
The uncertainty (= RSE / $\sqrt{N_{AGN}}$) of astrometric zero point shift is 0.072 mas.
In addition, the maximum and minimum of the parallaxes are 0.801 mas and -0.893 mas, respectively.
If we take into account the local zero-point shift of the {\it Gaia} parallax,
the real parallax of MBM 110 is 2.667 $\pm$ 0.095 mas (equivalently 375 $\pm$ 13 pc). However, the total error of the parallax
will be 0.448 mas if we take into account the RSE of the zero-point.
This result indicates that the random error of the {\it Gaia} DR2 parallaxes could be large, and therefore the {\it Gaia} DR2 parallax
of MBM 110 candidates may not be a decisive membership criterion.

\citet{lhb18} mentioned that a five-parameter solution is accepted only if at least six visibility periods (``visibility\_periods\_used''
in the {\it Gaia} archive) are used. The number of visibility periods of the MBM 110 candidate members is between 10 and 12,
which implies that the stars in MBM 110 are well observed during the {\it Gaia} nominal observation.
Another quantity of checking the quality of data derived from a multivariate data analysis is the reduced $\chi^2$ of the fit. 
\citet{ll18} used the square-root of 
the value as the unit weight error (UWE). The UWE of good data is about 1. However, the UWE distribution of the {\it Gaia}
DR2 parallax is a function of brightness as well as colour of the objects. \citet{ll18} provided a method of re-normalization and
relevant tables. We calculated the RUWE (re-normalized unit weight error) of candidate members of MBM 110. 
Among the stars in Table \ref{member}, 5 stars (WFI 496, 684, 1192, 9545, and 9818) have larger RUWEs [RUWE $\geq 1.4$
- see \citet{ll18} for the value]. These stars have a large error in the parallax, and hence are classified as probable members
even though their parallaxes are far different from those of member stars.

\subsection{{\it WISE} Mid-Infrared Data and Final Decision \label{wise}}

\begin{figure*}
\includegraphics[width=1.8\columnwidth]{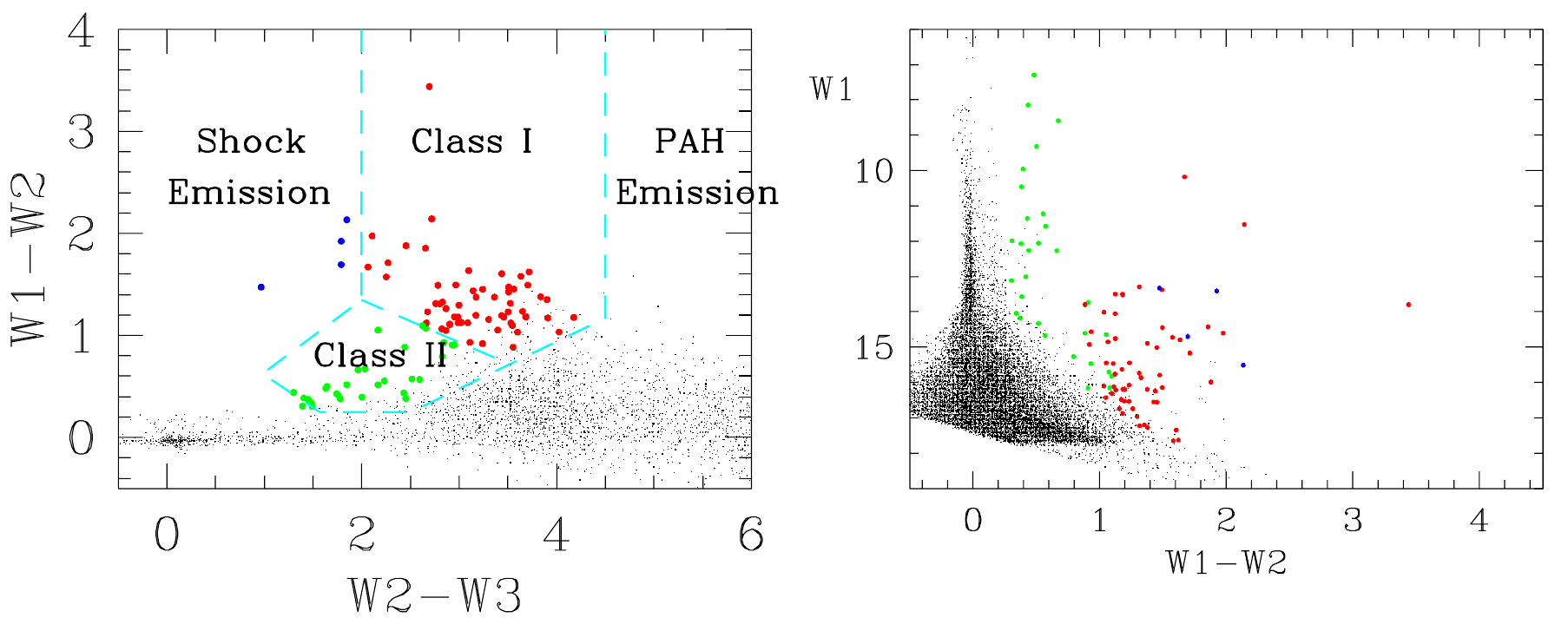}
\caption{{\it WISE} MIR two-colour diagram and colour-magnitude diagram.  YSO classification scheme in the left panel
is taken from \citet{fpss16}.}
\label{wise_ccd} 
\end{figure*}

For a comprehensive study of this SFR we have downloaded MIR WISE data from {\it VizieR}.
The YSO classification scheme
in the {\it WISE} TCD of \citet{fpss16} was adopted. In the WFI FOV, the number of Class I, Class II, Shock emission object,
and PAH emission objects is 51, 26, 4, and 0, respectively. As can be seen in the right panel of Figure \ref{wise_ccd},
the photometric errors increase rapidly for fainter objects ($W1 \gtrsim 15$). And so, if we limit the classification
to the bright objects ($W1 \leq 15$), the number is reduced to 22, 19, 3, and 0, respectively.
Among them, 1 Class I (WFI 9545), 13 Class II (see Table \ref{member}), and 1 Shock emission (WFI 6309) objects
have optical counterparts. Most Class I YSOs are  concentrated in the western part of the cloud, where many
Herbig-Haro objects can be found [see Figure \ref{map} (left)].

For the selection of members of MBM 110, we should take into account the photometric data, spectroscopic
characteristics, {\it WISE} MIR information, parallax, and proper motions. The first criterion for the membership
is parallax. From Figure \ref{fig_gaia}(a) we restrict our attention to the stars between $\tilde{\omega}$ = 2.40 -- 2.84 mas,
which covers most of the stars in the peak. And then we consider the other four criteria - proper motion vector,
H$\alpha$ emission from photometry, strong Li {\small I} $\lambda 6708\AA$ absorption and/or H$\alpha$ emission
in the spectrum, and MIR excess emission from {\it WISE} data. If a star meets at least two criteria from among
the 4 membership criteria, we classify the star as a member of MBM 110.
If a star meets two or more membership criteria mentioned above, but the parallax of a star is smaller or larger
than the range mentioned above we consider this star as a probable member of MBM 110. If a star meets the
membership criteria based on astrometry alone (parallax and proper motion) we consider the star as a possible
member of MBM 110. We cannot exclude the possibility that some young stars are in the foreground or background
of MBM 110, which is the member of the Orion OB1a association or the Orion-Eridanus superbubble.\\

\begin{table*}
 \caption{Summary of Membership Selection}
  \label{member}
   \begin{tabular}{clccccccccccl}
    \hline
     WFI &  other name & SSO & MAO & W(H$\alpha$) & W(Li) &
     SpT & WISE & $\tilde{\omega}$ (mas) & $\mu^{a}$ & RUWE$^{b}$ & Memb & Remark \\
    \hline
43 & Kiso A-0975 43 &  ...  &  ...  &  ...  &  ...  &  ...  & - & 3.035 
     & N & 1.29 & N & OEBM$^{c}$\\
496 & &  ...  &  ...  &  ...  &  ...  &  ...  & Class II & 4.685 
        & Y & {\bf 2.00} & P & large error in $\tilde{\omega}$\\
551 & Kiso A-0975 47 & H &  ...  & -63.4, -107.5 & 0.24 
       & K3 & Class II & 3.268 $\pm$ 0.025 & Y & 1.16 & N & OEBM \\
684 & & N &  ...  & 0.23 & 0.30 & K0 & - & 1.694 
        & Y & {\bf 8.10} & P &  large error in $\tilde{\omega}$\\
779 & & N &  ...  & -0.86, -1.30 & 0.57 $\pm$ 0.03 & K2 & - & 2.550 
       & Y & 1.00 & Y & \\
782 & & H &  ...  &  ...  &  ...  &  ...  & - & 1.347 
       & N & 1.30 & N & close double \\
842 & & N &  ...  & -4.83 & 0.06 & M2 & - & 6.442 
        & N & 1.25 & N & foreground star\\
881 & & N &  ...  & -4.58 & no abs & M2 & - & 4.160 
       & N & 1.09 & N & foreground star\\
899 & & N &  ...  & -79.2 &  ...  & M8 & & 2.674 
        & Y & 1.01 & Y & young BD member \\
904 & & N &  ...  &  ...  &  ...  &  ...  & - & 2.491 
        & Y & 0.95 & P & \\
955 & & N & N & -4.91 & 0.78 & M3 &- & 2.471 
        & Y & 1.05 & Y & \\
981 & HD 34835 & N &  ...  & 6.58 &  ...  & B6 & - &  2.578 
       & Y & 0.98 & Y & B5III/IV \\
983 & & N & N & -5.37 & 0.57 & M2 & - & 3.095 
        & N & 1.11 & P & OEBM\\
1001 & & N & N & -18.3, -4.85 & 0.36 $\pm$ 0.02 & G9/M3 & - & 2.624 
         & Y & 1.37 & Y & \\
1026 & & N & N &  ...  &  ...  &  ...  & - & 2.490 
         & Y & 1.03 & P & \\
1030 & & N & N & -3.84 & 0.39 & K2 & - & 2.584 
         & Y & 1.02 & Y & \\
1032 & & H &  ...  &  ...  &  ...  &  ...  & Class II & 2.770 
         & Y & 1.11 & Y & \\
1033 & & N & N & +0.72, +0.18 & 0.29, 0.18 & G5 & - & 2.620 
         & Y & 1.09 & Y & \\
1081 & StH$\alpha$ 37 & H & H & -166.8, -37.4 & 0.36 $\pm$ 0.01 & K1/5 & Class II & 2.493 
         & Y & 1.14 & Y &  \\
1129 & & N & N &  ...  &  ...  &  ...  & - & 2.693 
         & Y & 1.31 & P & \\
1132 & StH$\alpha$ 38& H & H & -109.2, -91.3 & 0.28, 0.14 & K4 & Class II & 2.614 
         & Y & 1.21 & Y & W(H$\beta$) = -31.4\\
1153 & & N & N &  ...  &  ...  &  ...  & Class II & 2.712 
         & Y & 0.99 & Y & \\
1161 & & N &  ...  &  ...  &  ...  &  ...  & - & 2.528 
        & Y & 1.01 & P & \\
1170 & HD 34890 & N &  ...  & 5.53 &  ...  & B9 & - & 2.751 
         & Y & 1.01 & Y & SB1 \\
1192 & StH$\alpha$ 39 & h & H & -20.3 & 0.44 & K1 &  Class II & -0.640 
         & N & {\bf 51.13} & P & large error in $\tilde{\omega}$ and $\mu$\\
1218 & & N &  ...  &  ...  &  ...  &  ...  & - & 2.834 
         & Y & 1.00 & P & \\
1235 & & N & N &  ...  &  ...  &  ...  & - & 2.814 
         & Y & 1.00 & P & \\
1240 & & H & H &  ...  &  ...  &  ...  & Class II & 2.962 
         & N & 0.90 & P & \\
1276 & & N &  ...  &  ...  &  ...  &  ...  & - & 2.621 
         & Y & 0.98 & P & \\
1283 & & N & h &  ...  &  ...  &  ...  & - & 3.183 
         & Y & 1.00 & P & OEBM\\
1314 & Kiso A-0975 56 & (H) & H &  ...  &  ...  &  ...  & Class II & 2.579 
         & Y & 1.11 & Y & \\
1327 & Kiso A-0975 57 & H & H & -40.8, -30.4 & 0.38, 0.90 & K7 & Class II & 2.577 
         & Y & 1.15 & Y & H$\gamma$ \& H$\delta$: stronger than H$\beta$ \\
1360 & & H &  ...  & -1.34 & no abs & K7 & - & 2.086 
         & N & 1.14 & N & \\
1372 & & N &  ...  &  ...  &  ...  &  ...  & - & 2.737 
         & Y & 1.09 & P & \\
1454 & & N &  ...  & +0.89 & 0.21 & M1 & - & 0.180 
         & N & 0.98 & P & large error in $\mu$\\
1459 & & N & N &  ...  &  ...  &  ...  & Class II & 2.877 
         & Y & 1.03 & Y & \\
1488 & & H & H &  ...  &  ...  &  ...  & Class II & 2.311 
         & N & 1.24 & P & OEBM\\
1522 & Kiso A-0975 59 & H & H  & -92.1 & 0.45 & K1 & Class II & 2.528 
         & Y & 1.06 & Y & \\
1707 & 1RXS J052125.6 & N &  ...  & +2.58 & +0.05 &  ...  & - & 2.751 
         & Y & 1.09 & Y & \\
5897 & & H &  ...  &  ...  &  ...  &  ...  & - & 0.091 
          & N & 1.11 & N & large error in $\tilde{\omega}$ and $\mu$\\
6309 & & H &  ...  &  ...  &  ...  &  ...  & Class II &  ...  %
          &  ...  & ... & P & no data in Gaia DR2\\
9545 & & & &  ...  &  ...  &  ...  & Class II &  ...  %
          &  ...  & {\bf 4.23} & P &  no data in Gaia DR2\\
9818 & & h &  ...  &  ...  &  ...  &  ...  & - & 1.543 
         & Y & {\bf 1.85} & N & large error in $\tilde{\omega}$ and $\mu$\\
\hline
\end{tabular}

$^{a}$: membership from proper motion, $^{b}$RUWE: renomalized unit weight error, $^{c}$OEBM: Orion-Eridanus bubble member?\\
\end{table*}

\section{PHYSICAL PROPERTIES}

\subsection{Adopted Parameters and Calibrations}

The mass and age of a star can be derived from the HRD with the help of stellar evolution models and
PMS evolution tracks. Various parameters and calibrations need to be adopted for the construction of the HRD.
The first parameters are those related to the reddening correction. HD 34385 is the only star which can be used 
to determine the reddening $E(B-V)$ and $R_V$. $E(B-V)$ = 0.11 and $R_V$ = 3.42
are adopted and applied to all member stars. The reddening in other colours is determined using
the relations between $R_V$ and colour excess ratio adopted in \citet{slb13}.
The distance of MBM 110 is based on {\it Gaia} DR2 parallaxes corrected for
the local zero-point shift from local {\it WISE} AGN samples - $\tilde{\omega}$ = 2.667 $\pm$ 0.095 mas
($d = 375 \pm 13 pc$).

We shortly describe the calibrations most relevant to this work. As most member stars in MBM 110 are low-mass stars,
the temperature and bolometric  correction (BC) scales for low-mass stars used here should  be described.
Although the surface gravity of low-mass PMS stars is intermediate between giants and MS, their photometric
characteristics, especially in $VRI$ passbands are very similar to MS stars \citep{llb04}. We use the temperature and BC scales of
\citet{pm13} (hereafter PM13) for MS stars.  Their colour-temperature relations are consistent 
with the theoretical colour-temperature
relations of BHAC of age 1 Gyr.  The difference in $\log T_{eff}$ from ($R-I$), ($V-I$), ($V-J$), ($V-H$), and ($V-Ks$) 
between PM13 and the 1 Gyr isochrone of BHAC is $+0.004 \pm 0.009$, $-0.000 \pm 0.004$, $+0.005 \pm 0.003$,
$+0.004 \pm 0.004$, and $+0.006 \pm 0.004$, respectively. And the difference between PM13 and the 5 Myr isochrone
of BHAC is slightly larger, and is $0.016 \pm 0.028$, $0.006 \pm 0.010$, $0.012 \pm 0.014$, $0.005 \pm 0.010$,
and $0.008 \pm 0.011$ for ($R-I$), ($V-I$), ($V-J$), ($V-H$), and ($V-Ks$), respectively.
We obtained the effective temperature and BC from PM13 using ($B-V$), ($R-I$), ($V-I$), ($V-J$), ($V-H$), and ($V-Ks$).
However, as $B$ could be affected by UV excess, $R$ by H$\alpha$ emission, and $Ks$ by IR excess, 
we applied different weights for different colors - weight = 0.0 for ($B-V$),
0.5 for ($R-I$) and ($V-Ks$), and 1.0 for the others. We adopted the colour-temperature relations for MS stars, but the stars
in MBM 110 are young stars. We should check the reliability of the adopted calibrations. Although PM13 published
the colour-temperature relations for the stars with age 5 -- 30 Myr, their stars may be affected by UV excess due to mass
accretion, the uncertainty in reddening correction, the existence of circumstellar disks, etc. And therefore, it is better to check
the reliability using the theoretical relations of BHAC.
The temperature from colours differs by $+0.009 \pm 0.011$, $+0.006 \pm 0.007$,
$+0.004 \pm 0.005$, $+0.009 \pm 0.006$, and $+0.007 \pm 0.005$ in logarithmic scale for ($R-I$), ($V-I$), ($V-J$),
($V-H$), and ($V-Ks$), respectively. And the average difference in BC at a given temperature is only $-0.054 \pm 0.087$.
And therefore we can conclude that the adotpted relations do not cause a serious problem in temperature and BC estimates.
$I$ or $V$ magnitude is used in the calculation of $M_{bol}$.

The classical T Tau stars ($W_{H\alpha} \gtrsim 10\AA$) in the observed FOV (WFI 551, 1081, 1132, 1314, 1327, \& 1522)
showed variabilities in magnitudes and colours. We calculated the effective temperatures, BC's, masses and ages using
the individual data, and the average values were used in the HRD and initial mass function (IMF) calculation.

\subsection{Age  and Mass \label{age}}

\begin{figure*}
   \includegraphics[width=1.8\columnwidth]{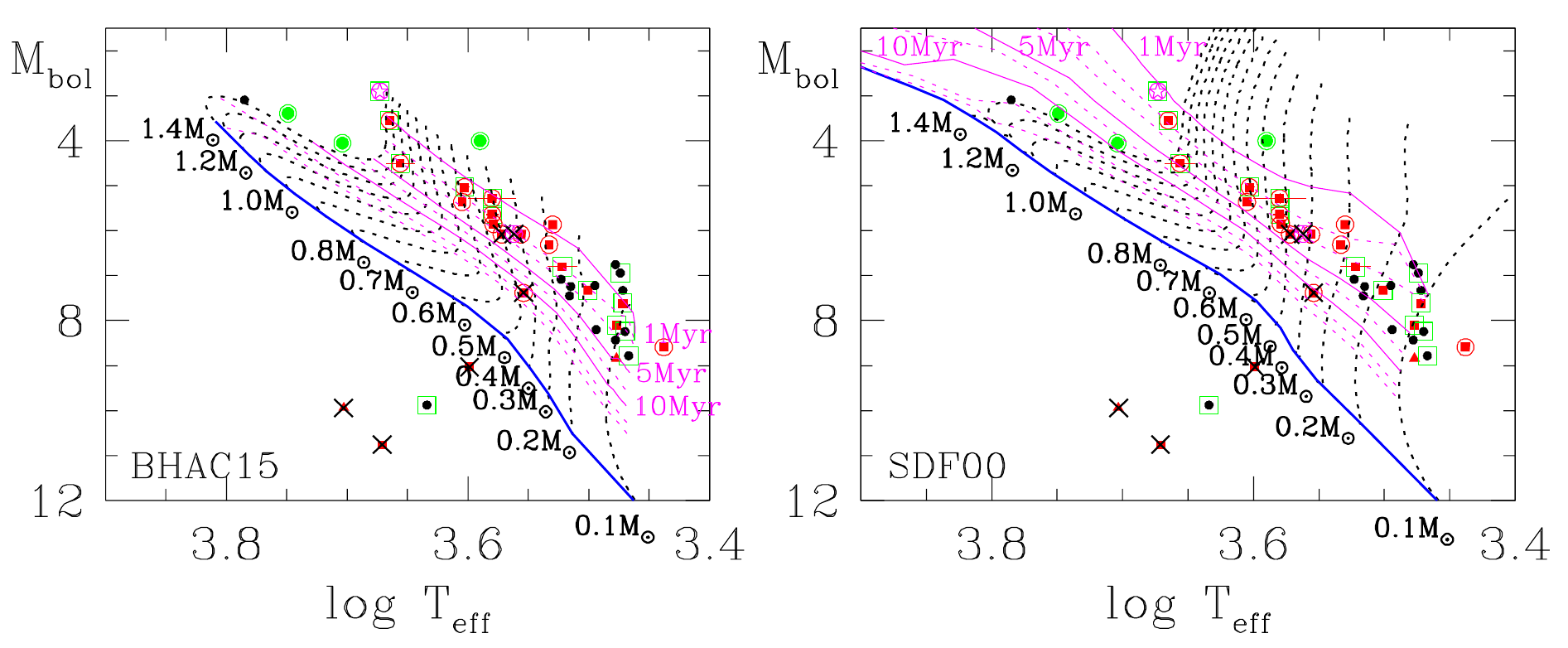}
     \caption{Hertzsprung-Russell diagram of MBM 110 assuming that the stars in Table \ref{member} are at the same distance
      of $375 pc$. The thick blue solid line in each diagram is the ZAMS relation.
      The magenta solid lines represent the isochrones of age 1, 5, and 10 Myr interpolated from the PMS evolution tracks,
      while the magenta dashed lines are the isochrones of ages 2, 3, 7, 15, and 20 Myr. The dotted lines with mass to
      the left or right are the PMS evolution tracks for the mass. The other symbols are the same as in Figure \ref{m4k_cmd}
     except for black crosses which represent non-member stars in Table \ref{member}. The left panel is based on
     the PMS evolution models by \citet{bhac15}, and the right panel is based on the PMS evolution models by \citet{sdf00}. }
  \label{fig_hrd} 
\end{figure*}

The HRD of MBM 110 is presented in Figure \ref{fig_hrd}.
We added error bars for 6 classical T Tau stars due to variability.
Most low-mass young stars in MBM 110 locate between the isochrone of age 1 Myr and 5 Myr.
However, several massive stars ($m \gtrsim 1 M_\odot$) seem to have older ages, which is a well-known feature of
PMS evolution models \citep{sbl97,sbc04}. The mean age of 9 low-mass PMS members ($m$ = 1.0 -- 0.1 $M_\odot$)
is 1.9 ($\pm$ 0.9) Myr from the PMS evolution models by BHAC and 3.1 ($\pm$ 1.2) Myr from \citet{sdf00} (hereafter SDF).

The mass of individual stars can be determined in the HRD by interpolating the evolutionary tracks. Two evolutionary models
give very similar masses for a given star. The mass scale for member stars between two PMS models is

$$ \log m_{BHAC} = 1.018 (\pm 0.058) \log m_{SDF} - 0.012 (\pm 0.032). $$

\noindent
Because of the mass coverage, as well as consistency with our previous studies, we use the masses from
SDF models. The most massive star in MBM 110 (WFI 981 = HD 34835) is just about $3.9 M_\odot$.
And the total stellar mass of MBM 110 is about 16.1 $M_\odot$ for member stars and 23.1 $M_\odot$
including probable members. 

\begin{figure}
   \includegraphics[width=\columnwidth]{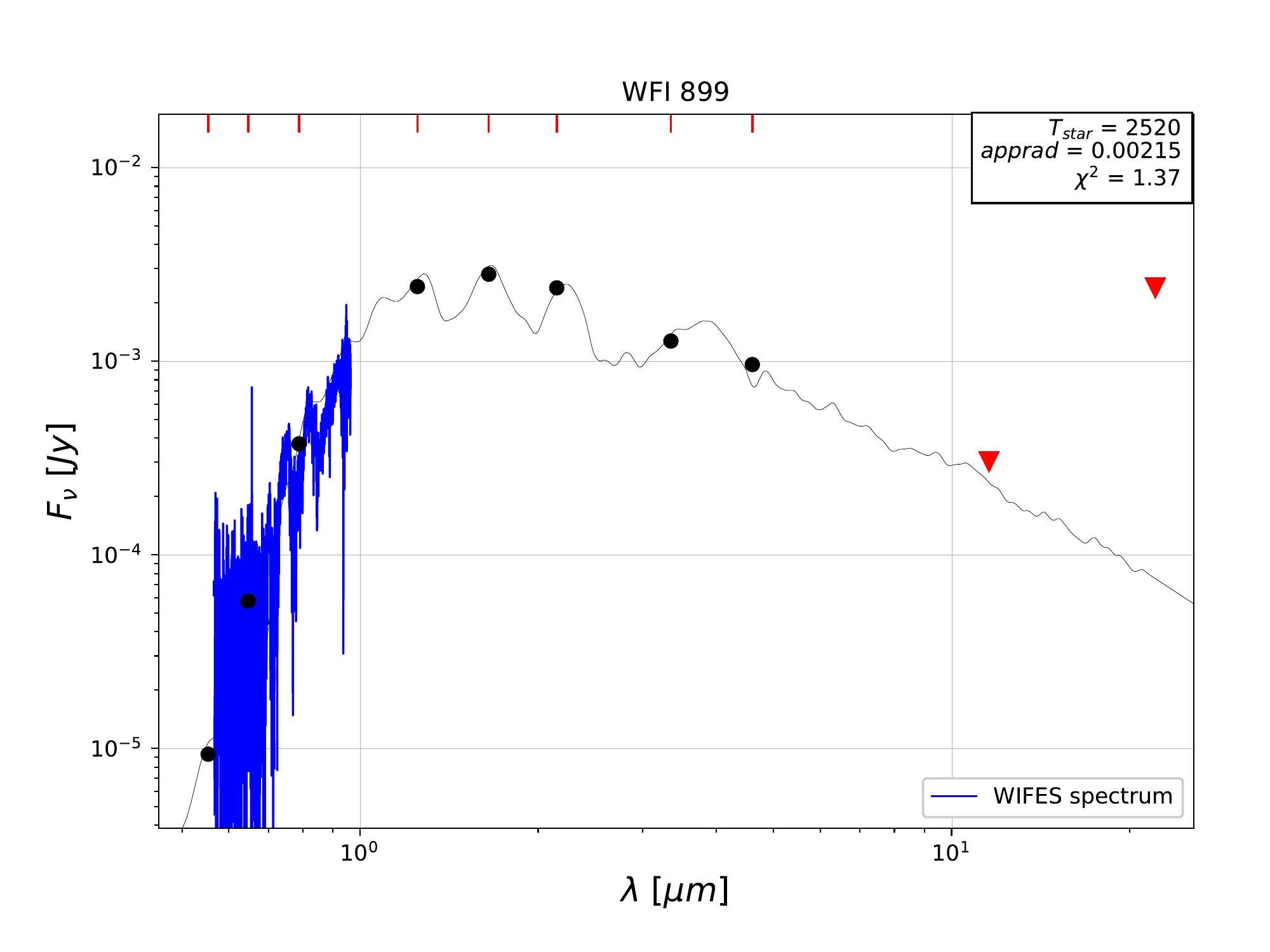}
     \caption{The spectral energy distribution (SED) of WFI 899. Filled circles represent photometric data from SSO WFI, 2MASS, 
     and WISE. Two reversed triangles denote the upper limits from WISE. The shaded area represents the spectrum of WFI 899.
     The solid line indicates the best fit SED of WFI 899 using the updated NextGen model (\citealt{hab99} and priv. comm.). }
   \label{sed}
\end{figure}

The effective temperature and $M_{bol}$ of the young brown dwarf member WFI 899 is 2740K and +8.59 mag, respectively.
The star is redder and brighter than the 0.5 Myr isochrone of BHAC, and its mass is estimated to be about 46 $M_J$. 
The spectral energy distribution (SED) of WFI 899 is shown in Figure \ref{sed}. The SED fitting results give
$T_{eff} =$ 2520K and  the apparent radius $\phi$ = 2.15 mas. The radius $0.81 R_\odot$ (at 385 pc) is too large
as a young brown dwarf. The calculated luminosity is too bright and it corresponds to about $100 M_J$ object.
If WFI 899 is a foreground young brown dwarf at about 120 pc, then $\log L/L_\odot = -2.61$, which corresponds 
to a $20 M_J$ brown dwarf. The discrepancy between temperature and luminosity is also disentangled.
However, the distance from {\it Gaia} DR2 is consistent with that of MBM 110. This discrepancy may be
resolved in future data releases from the {\it Gaia} astrometric mission.

If the {\it Gaia} parallax of WFI 551 is correct,
the star is a young star in the foreground, and therefore a member of the Orion-Eridanus superbubble.
In addition, if WFI 551 is a low-mass foreground PMS star, it is 0.48 mag fainter than the members
of MBM 110 in Figure \ref{fig_hrd}, and its age is about 5 Myr from SDF or 3.5 Myr from BHAC.
However, it is very difficult to accept that active accretion persists for 5 Myr
and although the possibility of it being a PMS star with a nearly edge-on disk cannot be ruled out,
it is very difficult to explain the inverse P-Cyg profile of the star.

\begin{figure}
   \includegraphics[width=\columnwidth]{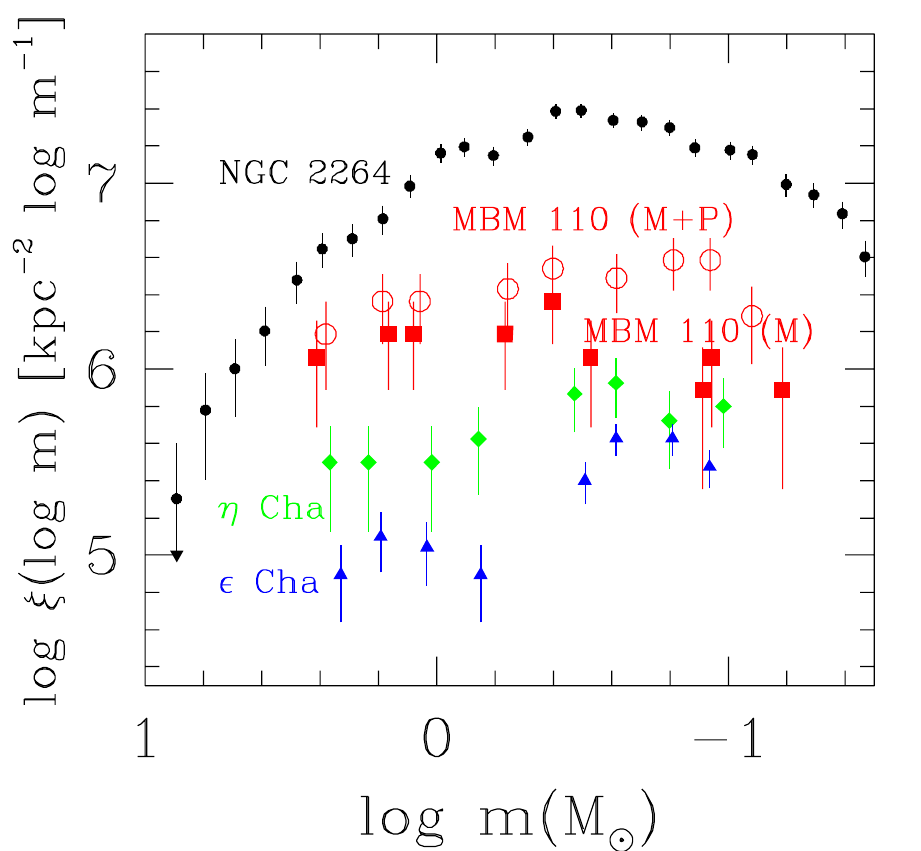}
     \caption{The IMF of MBM 110 (red). Red square and red circle represents
     the IMF of member stars only and that of members and probable members, respectively.
     For comparison, we also present the IMF of several SFRs
     - NGC 2264 (black), $\eta$ Cha (green), and $\epsilon$ Cha (blue). As the number of member stars is very small,
     we calculate the initial mass function $\xi (log ~m)$ in a bin size of $\Delta \log ~m =0.4$ for MBM 110, $\eta$ Cha, and
     $\epsilon$ Cha. To avoid the binning effect, we calculate the IMF by shifting the bin by 0.2 in $\log m$. However, the IMF
     of NGC 2264 is for $\Delta \log ~m =0.2$. }
   \label{fig_imf}
\end{figure}

Figure \ref{fig_imf} shows the initial mass function (IMF) of MBM 110 as well as that of three nearby star-forming regions (SFRs).
Our spectroscopic census is complete for $I \lesssim$ 14.5 mag, which is about $0.4 M_\odot$ for the 5 Myr
isochrone. However, our photometric search for H$\alpha$ emission stars from H$\alpha$ photometry is complete
down to $I \approx$ 17 mag, equivalently $0.08 M_\odot$. In addition, the completeness of {\it Gaia} astrometric data 
is similar to our photometry. And therefore, the completeness of the IMF is about $0.1 M_\odot$.
The membership selection of $\eta$ Cha and $\epsilon$ Cha is described in the Appendix.
As the number of member stars is very small, we use a larger bin size of $\Delta \log ~m =0.4$.
We calculate the number of stars and mean mass in a logarithmic mass scale for each bin.
The IMF of NGC 2264 is obtained from \citet{sb10}. Most young open clusters \citep{sb10,sbc17} or field stars in
the Solar neighborhood \citep{Cha03} show a peak in the IMF at $\log m = -0.3$ -- $-0.7$.
However, the IMF of MBM 110 shows a peak at $\log m \approx 0.0$ for the IMF of members or a flat IMF in
$\log m = 0.0$ -- $-1.0$ for the IMF of members and probable members. The IMF of $\eta$ Cha is very similar
to that of MBM 110. However, the IMF of $\epsilon$ Cha is increasing down to $\log m \approx -0.8$.
The surface density of MBM 110 is about 10 times higher than that of $\eta$ Cha or $\epsilon$ Cha, mainly
due to the adopted radius of each SFR ($r = 1.^\circ 6$ for $\eta$ Cha and $4.^\circ 0$ for $\epsilon$ Cha).
The mass of the most massive star in MBM 110, $\eta$ Cha, and $\epsilon$ Cha is at most between 3.0 -- 4.0$M_\odot$.

\subsection{Star Formation Efficiency}

To determine the star formation efficiency of MBM 110, the mass of the molecular cloud should be calculated.
The mass of the molecular cloud MBM 110 is estimated 
from the CO luminosity [W(CO)], angular size of the source, and 
the well-determined conversion factor ($N_{\rm H_{2}} = 2.0 \times 10^{20}
\rm W(CO) ~cm^{-2} $ - \citealt{bwl13}). Because the angular size of MBM 110 was not mentioned
in \citet{mbm85}, we approximated the shape of the cometary cloud as a triangle in the {\it Akari} MIR image.
The projected area of the cloud (about $ 0.72 \rm deg^2$)  is equivalently $ 2.92 \times 10^{38} \rm cm^2$ at $d = 375 pc$.
However, the calculated W(CO) is far smaller than that presented by \citet{mmm86} [W(CO) = 14.5 $\rm K ~km ~s^{-1} $]
of MBM 110 (Cloud 14 in \citealt{mmm86}).
Considering the size of the MBM 110 cloud as determined from the $E(B-V)$ map of \citet{sfd98},  the larger beam ($8.'4$) used by
\citet{mmm86} in their mapping observation versus the single pointing observation with a larger telescope (e.g., smaller beam size) 
of \citet{mbm85}, the CO measurement of \citet{mmm86} is likely to give a better estimate of W(CO). The mass of the MBM 110 cloud
is about $1680 M_\odot$. The minimum value of the star formation efficiency of MBM 110 is 1.0\%. If we take into account
the mass of probable members, the efficiency is about 1.4\%. However, this value is a minimum value because many Class I objects
seen in the {\it WISE} image are embedded in the western part of the cloud.

\section{DISCUSSION}

\subsection{Triggered Star Formation versus Spontaneous Star Formation}

Recently \citet{kcs18} studied young stars in the Orion Star Forming Complex using {\it Gaia} DR2 and APOGEE-2 survey data.
They analyzed data for more than 10,000 stars in 6-dimensional space and identified 190 groups in 5 larger structure.
MBM 110 belongs to their Orion D - the oldest ($\sim$ 5 -- 8 Myr) group in the Orion SFR which comprises Orion OB1a and OB1b.
However, the age of MBM 110 is about 4 Myr, which is younger than IC 2118 (7.9 $\pm$ 2.5 Myr) and similar to LDN 1616
(5.0 $\pm$ 1.7 Myr) \citep{kcs18} or Ori OB1c group \citep{Bal08}.
And therefore, were the star formation in MBM 110 triggered by massive stars in the Orion OB association,
the massive stars in Ori OB1a or 1b subgroup (Orion D of \citealt{kcs18}) might be the triggers.

\citet{lc07} summarized the three distinct characteristics of triggered and spontaneous star formation in their Table 7
- signature of sequential star formation, spatial distribution of young stars and the location of ionization front, and
the star formation efficiency. If the star formation in MBM 110 was triggered by massive stars in the Orion OB association,
we can expect a systematic age variation along right ascension. However, we could not find any systematic variation of stellar
age in either direction. The star formation efficiency is also very low.
In addition, if the radiation-driven implosion model is operating in MBM 110, we could not find  young stars far behind 
the bright rim. However, as shown in Figure \ref{map}, many Class I YSOs and related objects such as
Herbig-Haro objects are found in LDN 1634. These facts for MBM 110 are not consistent with the triggered star formation scenario
summarized by \citet{lc07}.

\citet{dvns02} observed several molecular transition lines in $mm$ and $sub-mm$ wavelength
in order to find direct evidence for the role of massive stars in the bright-rimmed clouds including L1634 [the 16th object
in the \citet{sfo91} catalog]. However, they could, in most cases, not see any direct evidence of triggering in these sources.
These facts imply that spontaneous star formation may be operating in MBM 110.

\subsection{Star Formation with No Massive Stars}

The nearby Tau-Aur SFR is a well-known site of low-mass star formation \citep{kgw08}.
This kind of small SFRs seems to be ubiquitous. There are many small SFRs in the outskirt of the Orion nebula \citep{acl08}
as well as many young star groups in the Solar neighbourhood \citep{zs04}.
\citet{chs00} found that about 39\% of the cluster population, identified in the K' images of 32 IRAS point sources
distributed in the Cas OB6 association, are embedded in small clouds located as far as 100 pc from the W3/W4/W5 region.
They speculated that these small clouds are fragments of a cloud complex dispersed by previous episodes of massive star formation.
MBM 110 may be a similar object in the Orion-Eridanus superbubble. In this section,
we estimate the contribution of such low-mass SFRs to the global mass function.

\begin{table}
 \caption{Mass budget of star forming groups}
  \label{tab_mmax}
   \begin{tabular}{cccccc} 
    \hline
     $m_{\rm max}$ &  \# of & M$_{\rm ecl}^{\rm med}$ & $d_{\rm comp}$ & Surface density & M$_{\rm total}$ \\
     (M$_\odot$) & SFRs & (M$_\odot$)  &  (kpc) & (kpc$^{-2}$) &  (M$_\odot$ kpc$^{-2}$) \\
    \hline
$<$ 2 & 6 & 8 & 0.2 & 39.8 & 318\\
2 -- 4 & 20 & 28 & 0.2 & 71.6 & 2005\\
4 -- 8 & 10 & 59 & 0.5 & 5.1 & 301 \\
8 -- 16 & 34 & 108 & 1.0 & 3.2 & 346 \\
16 -- 32 & 29 & 949 & 3.0 & 0.95 & 902 \\
32 -- 63 & 25 & 3573 & 3.0 & 0.46 & 1644 \\
65 -- 150 & 14 & 9258 & 3.0 & 0.18 & 1637 \\
\hline
\end{tabular}
\end{table}

\begin{figure}
   \includegraphics[width=\columnwidth]{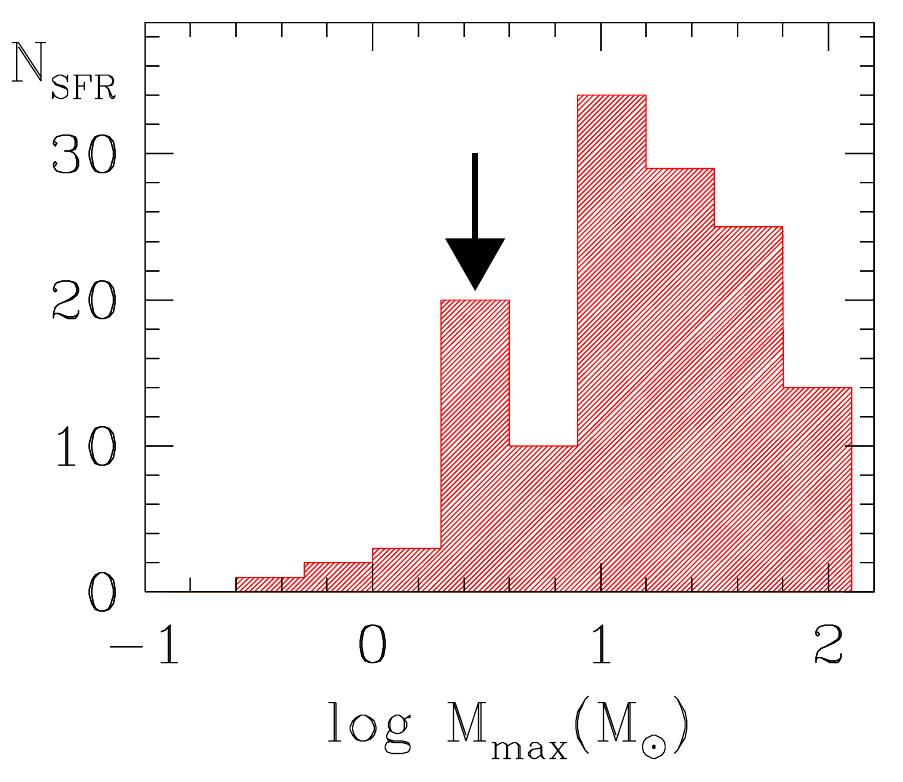}
     \caption{The maximum stellar mass $m_{max} (M_\odot)$ distribution of SFRs. The arrow indicates
     a distinct peak at $m_{max} = 2$ -- $4 M_\odot$.}
   \label{fig_mmax}
\end{figure}

Figure \ref{fig_mmax} shows
the distribution of mass of the most massive star [$m_{\rm max}$ (M$_\odot$)] in various SFRs from \citet{wkpa13}.
As marked in the figure, there is a distinct peak at $m_{max} = 2$ -- $4 M_\odot$ with a spectral type of B5 or later.
The mini-clusters $\eta$ Cha, $\epsilon$ Cha as well as MBM 110 belong to this category.
These small SFRs are faint, and therefore very difficult to recognize at a large distance.
We investigate the characteristics of these small SFRs using the data in Appendix of \citet{wkpa13}.
The cluster mass (M$_{\rm ecl}$ in \citealt{wkpa13}) of this type of SFR  is between 13 -- 126 $M_\odot$ and
its median value is 28 $M_\odot$, which is similar to the total stellar mass of MBM 110.
The distances of the SFRs in the bin are mostly $d \lesssim 200 pc$.\footnote{9 SFRs are closer than $200 pc$ and
7 SFRs are farther than $2 kpc$. However, the $m_{max}$ of the latter case may be very uncertain.} And therefore,
it is reasonable to assume that the data for these small SFRs are complete for $d \lesssim 200 pc$.

On the other hand, the highest peak in Figure \ref{fig_mmax} is for the clusters of $m_{\rm max}$ (M$_\odot$) = 8 -- 16 M$_\odot$,
which is equivalent to spectral type of O9V -- B3V. The median M$_{\rm ecl}$ of these SFRs is about 108 M$_\odot$.
The open clusters belonging to this group are bright enough to be observable even in the neighboring spiral arms.
The distance distribution of these clusters shows several peaks - $d \lesssim 1 kpc$ and between $d = 2$ -- $2.6 kpc$.
The first peak is the young open clusters in the local Orion spur, and those in the second peak are those in the Sgr-Car arm
or in the Per arm. The census of these clusters may be complete for $d \lesssim 1 kpc$. The number of
these clusters within $d = 1 kpc$ is 10, and therefore the surface density of this group is 3.2 $kpc^{-2}$.

Table \ref{tab_mmax} summarises the result from the analysis of young SFR data in \citet{wkpa13}. The first column shows
the range of $m_{\rm max}$ of each bin (bin size:  $\Delta \log m \approx 0.3$). The second and third column represents,
respectively, the number of SFRs and median M$_{\rm ecl}$ of each bin.
The fourth column is the distance of completeness for each group ($d_{comp}$), and the fifth column is the surface number
density of each group using the number of SFRs within $d_{comp}$. The last column shows the mass
contribution of each group to the Galactic disk within $1 kpc$. Interestingly, the $m_{max} = 2$ -- $4 M_\odot$ SFRs
contribute about 5.8 times that of the $m_{max} = 6$ -- $16 M_\odot$ SFRs. In addition, the contribution of small SFRs
is about 30\% of the total young stellar mass in the disk, and is the dominant component compared to any other groups in
Table \ref{tab_mmax} and Figure \ref{fig_mmax}. As the  M$_{ecl}$ of larger SFRs are suspected to be overestimated
\citep{sbc17}, the contribution of these small SFRs may be larger than this. The implication of this result seems to be
profound because
these small SFRs are easily missed from observation due to their faintness. In addition, the mass of the most massive star
in the small SFRs is not enough to contribute to the H$\alpha$ luminosity of galaxies. And therefore the star formation rate
calculated from the H$\alpha$ luminosity may underestimate the actual star formation rate.

\section{Summary and Conclusions}

We presented the optical photometric and spectroscopic data for stars in the high Galactic latitude molecular
cloud MBM 110 which is also known as the dark nebula L1634. For a comprehensive study of MBM 110 we analyzed
{\it WISE} MIR data as well as {\it Gaia} astrometric data. The membership of stars in the observed field was
critically evaluated for each star. A total of 17 members and 18 probable members were selected from this procedure.
The results obtained from this study are as follows.

(1) The mean parallax of MBM 110 is 2.667 $\pm$ 0.095 mas (equivalently $d = 375 \pm 13 pc$) 
determined from the {\it Gaia} DR2 data. The age of young stars in MBM 110 is about 1.9 Myr from \citet{bhac15}
or 3.1 Myr from
\citet{sdf00}. The high mass truncation in the IMF occurs at about $4 M_\odot$.

(2) The star formation efficiency of MBM 110 is very low. The minimum value is 1.0\%. The efficiency
will be 1.4\% if probable members are included.

(3) The young stars in MBM 110 do no show any systematic variation in age along the right ascension direction.
And star formation efficiency is very low.  In addition, many Class I YSOs and related objects can be found
far away from the ionization front, which implies that star formation in MBM 110 was not triggered by the radiation-driven
implosion process. These facts are not consistent with the signatures of triggered star formation summarized by
\citet{lc07}.

(4) The total stellar mass of MBM 110 is very small and is comparable to that of other small SFRs in the Solar neighborhood.
It seems that although such small SFRs are very faint they are also very numerous. We analyse the mass budget of young SFR groups,
and argue for the importance of small SFRs in the global star formation rate.
Because these small SFRs are easily missed from observation due
to their faintness, and because the mass of the most massive star in
the small SFRs is insufficient to contribute to the H$\alpha$  luminosity of
galaxies, the star formation rate calculated from the H$\alpha$ luminosity 
very likely underestimates the actual star formation rate.

\addcontentsline{toc}{section}{Acknowledgements}

This work has made use of data from the European Space Agency (ESA) mission
{\it Gaia} (\url{http://www.cosmos.esa.int/gaia}),
processed by the {\it Gaia} Data Processing and Analysis Consortium (DPAC,
\url{http://www.cosmos.esa.int/web/gaia/dpac/consortium}). Funding
for the DPAC has been provided by national institutions, in particular
the institutions participating in the {\it Gaia} Multilateral Agreement.
H.S. acknowledges the support of the National Research Foundation of Korea (Grant No. NRF- 2015R1D1A1A01058444
and NRF-2019R1A2C1009475).








\appendix
\section{Membership Selection of the $\eta$ Cha Mini-Cluster}
The {\it Gaia} DR2 parallax and proper motion data for 14 known members of the $\eta$ Cha mini-cluster \citep{llb04}
were downloaded from
{\it VizieR}, and the average value and standard deviation of $\tilde{\omega}$, $\mu_{\alpha}$, and $\mu_\delta$ were calculated.
And then we searched for the probable members around $\eta$ Cha within a $2^\circ$ radius and within 5 times the standard
deviation around the average ($\tilde{\omega}$ = 8.95 -- 11.42 mas, $\mu_{\alpha}$ = -35.5 -- -23.6 mas/yr, and $\mu_\delta$
= +24.3 -- +29.8 mas/yr). From this procedure we selected 9 candidates. One of them ({\it Gaia} DR2 5196802989555695104)
is fainter than the PMS sequence of $\eta$ Cha, and so the star was deleted from the list.

The characteristics of the final 21 members of the $\eta$ Cha mini-cluster are $\tilde{\omega}$ = 10.164 $\pm$ 0.036 mas,
$\mu_\alpha$ = -29.53 $\pm$ 1.90 mas/yr, $\mu_\delta$ = +27.09 $\pm$, and $v_r$ = +14.05 $\pm$ 4.55 km/s. And the radius
of $\eta$ Cha ($r = 1.^\circ 6$, is equivalently 2.75 pc at $d$ = 98.3 pc) is determined from the maximum distance of members
from $\eta$ Cha.

\section{Membership Selection of the $\epsilon$ Cha Mini-Cluster}
From Fig. 2 of \citet{mlb13} we selected 77 candidate $\epsilon$ Cha cluster stars within $4^\circ$ from ($\alpha$, $\delta$)
= ($12^h 12.^m 5$, $-77^\circ$) with $\tilde{\omega}$ = 9.12 -- 10.57 mas, $\mu_\alpha$ = -50.2 -- -31.2 mas/yr, and $\mu_\delta$
= -30.7 -- +17.4 mas/yr. Among them 11 stars were rejected in the CMD, 6 stars were deleted due to a larger difference
from the mean and standard deviation of the proper motion vector. However, 4 stars ($\epsilon$ Cha AB, $\epsilon$ Cha 17,
$\epsilon$ Cha 36, \& $\epsilon$ Cha 24) that were rejected due to a large error or no data in {\it Gaia} DR2 measurements
were recovered.

The astrometric characteristics of 64 members of $\epsilon$ Cha are $\tilde{\omega}$ = 9.65 $\pm$ 0.74 mas, $\mu_\alpha$ =
-40.53 $\pm$ 1.75 mas/yr, and $\mu_\delta$ = -6.97 $\pm$ 3.88 mas/yr. However, \citet{mlb13} considered that
the size of $\epsilon$ Cha is much larger than the searching radius used here. In that case the IMF of $\epsilon$ Cha
in Figure \ref{fig_imf} will be lower by about 1 dex.


\bsp	
\label{lastpage}
\end{document}